\documentclass[sigconf]{acmart}
\newcommand{\modelname}{\textsf{DINS}\xspace}
\usepackage{xspace}

\makeatletter
\DeclareRobustCommand\onedot{\futurelet\@let@token\@onedot}
\def\@onedot{\ifx\@let@token.\else.\null\fi\xspace}

\def\eg{\emph{e.g}\onedot} 
\def\ie{\emph{i.e}\onedot}

\makeatother

\AtBeginDocument{%
  \providecommand\BibTeX{{%
    \normalfont B\kern-0.5em{\scshape i\kern-0.25em b}\kern-0.8em\TeX}}}

\copyrightyear{2023}
\acmYear{2023}
\setcopyright{acmlicensed}\acmConference[CIKM '23]{Proceedings of the 32nd ACM International Conference on Information and Knowledge Management}{October 21--25, 2023}{Birmingham, United Kingdom}
\acmBooktitle{Proceedings of the 32nd ACM International Conference on Information and Knowledge Management (CIKM '23), October 21--25, 2023, Birmingham, United Kingdom}
\acmPrice{15.00}
\acmDOI{10.1145/3583780.3614845}
\acmISBN{979-8-4007-0124-5/23/10}

\usepackage{multirow,makecell}
\usepackage{enumitem}
\usepackage{xspace}
\usepackage{color}
\usepackage{amsmath}
\usepackage{subcaption}
\usepackage{algorithm}
\usepackage{algorithmic}
\usepackage{booktabs}

\DeclareMathOperator*{\argmax}{arg\,max}

\usepackage[normalem]{ulem}
\useunder{\uline}{\ul}{}
\begin{document}

\title{Dimension Independent Mixup for Hard Negative Sample in Collaborative Filtering}

\author{Xi Wu}\authornote{Both authors contributed equally to this work.}
\affiliation{%
  \institution{YanShan University}
  \city{Qinhuangdao}
  \country{China}}
\email{wuxi@stumail.ysu.edu.cn}

\author{Liangwei Yang}\authornotemark[1]
\affiliation{%
  \institution{University of Illinois Chicago}
  \city{Chicago}
  \country{USA}}
\email{lyang84@uic.edu}

\author{Jibing Gong}\authornote{Corresponding author}
\affiliation{%
  \institution{YanShan University}
  \city{Qinhuangdao}
  \country{China}}
\email{gongjibing@ysu.edu.cn}

\author{Chao Zhou}
\affiliation{%
  \institution{YanShan University}
  \city{Qinhuangdao}
  \country{China}}
\email{chaozhou1999@gmail.com}

\author{Tianyu Lin}
\affiliation{%
  \institution{YanShan University}
  \city{Qinhuangdao}
  \country{China}}
\email{lintianyu@stumail.ysu.edu.cn}

\author{Xiaolong Liu}
\email{xliu262@uic.edu}
\author{Philip S. Yu}
\email{psyu@uic.edu}
\affiliation{%
  \institution{University of Illinois at Chicago}
  \city{Chicago}
  \country{USA}}

% \renewcommand{\shortauthors}{Liangwei et al.}
\renewcommand{\shortauthors}{Xi Wu et al.}

\begin{abstract}
Collaborative filtering (CF) is a widely employed technique that predicts user preferences based on past interactions. Negative sampling plays a vital role in training CF-based models with implicit feedback. In this paper, we propose a novel perspective based on the sampling area to revisit existing sampling methods. We point out that current sampling methods mainly focus on Point-wise or Line-wise sampling, lacking flexibility and leaving a significant portion of the hard sampling area un-explored. To address this limitation, we propose Dimension Independent Mixup for Hard Negative Sampling (\modelname), which is the first Area-wise sampling method for training CF-based models. \modelname comprises three modules: Hard Boundary Definition, Dimension Independent Mixup, and Multi-hop Pooling.
Experiments with real-world datasets on both matrix factorization and graph-based models demonstrate that \modelname outperforms other negative sampling methods, establishing its effectiveness and superiority. Our work contributes a new perspective, introduces Area-wise sampling, and presents \modelname as a novel approach that achieves state-of-the-art performance for negative sampling. Our implementations are available in PyTorch\footnote{\url{https://github.com/Wu-Xi/DINS}}.
\end{abstract}





\maketitle

\section{Introduction}
% \footnotetext[1]{Both authors contributed equally to this work.}

% Importance of RecSys

\begin{figure}
    \begin{center}
    \includegraphics[width=0.4\textwidth]{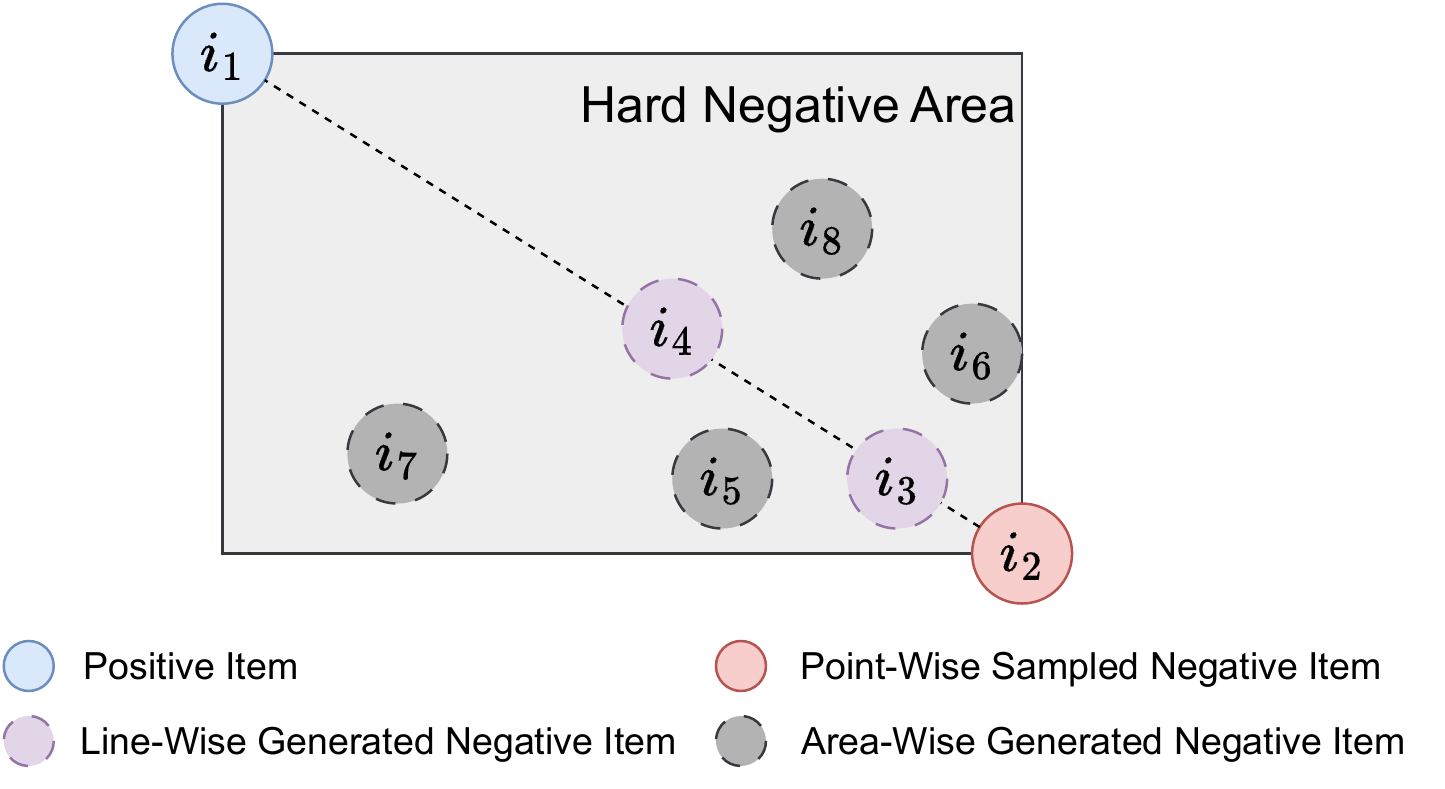}
    \end{center}
    \caption{Illustration of items sampled by Point/Line/Area-wise sampling methods.}
    \label{fig:illustration}
\end{figure}

In the contemporary era of voluminous data~\cite{bigdata}, individuals are inundated with an incessant influx of content generated by the internet. To address the issue of information overload, Recommender Systems (RecSys) are employed to assist users in locating the most relevant information and are increasingly pivotal in online services such as news feed~\cite{npa}, music suggestion~\cite{musicrec}, and online shopping~\cite{shoprec}. Collaborative filtering (CF)~\cite{cfrec}, a highly effective method that predicts a user's preference based on their past interactions, is widely employed. The latest CF-based models~\cite{He2020lightgcn,Wang2019NGCF} incorporate historical interactions into condensed user/item vectors and predict a user's preference for each item based on the dot product of the corresponding vectors. These models have garnered significant research attention and have been demonstrated to be effective in a range of application contexts~\cite{pinsage,yang2022large}.

% Importance Negative sample in training RecSys
Negative sampling~\cite{negative_sampling_origin,DNS,chen2023revisiting,liu2021dense} is a key technique when training CF-based models with implicit feedback~\cite{rendle2014improving}, which is inferred from user behavior, such as clicks, views, and purchases, rather than being explicitly provided by the user. Since implicit feedback is prevalent in most online platforms, it is frequently utilized in training RecSys~\cite{He2020lightgcn,Huang2021MixGCF,pinsage}. These behaviors only signify a user's positive feedback, necessitating the integration of a negative sampling module to provide negative feedback. The training process of CF-based models involves the differentiation between positive and negative examples, enhancing its ability to suggest items of interest to the user. The negative sampling approach has a significant impact on the ultimate performance of CF-based models~\cite{DNS,Huang2021MixGCF,Wang2017IRGAN:,SRNS,SoftBPR}. 

% How currently methods works, and their weakness, illustrated with motivation figure as we discussed

For each observed user-item interaction, the negative sampling module samples one or multiple negative items~\cite{chen2017sampling}. By introducing the concept of sampling area analysis, we offer a fresh perspective for understanding and categorizing these methods. In accordance with the proposed framework, this paper explores the sampling of negative items in relation to observed user-item interactions. As shown in Figure~\ref{fig:illustration}, $i_1$ is the observed positive item. Notably, the sampled negative items, denoted as $i_2$ to $i_8$, are obtained through diverse methodologies, which can be further categorized into Point/Line/Area-wise negative sampling methods.

\begin{enumerate}[leftmargin=*]
    \item The \textbf{Point-wise} sampling method ($i_2$) selects a particular item from the candidate item set using a predetermined sample distribution. In terms of embedding space interpretation, these methods involve selecting a single point from a predefined set of candidate points, each representing a potential candidate item. This category encompasses a majority of existing methods~\cite{Rendle2012BPR, DNS, Wang2017IRGAN:,AdvIR,ding2020simplify,RecNS}, as they employ sampling techniques within the discrete space. 
    \item The \textbf{Line-wise} sampling method ($i_3$, $i_4$) involves selecting a pseudo-negative item positioned along a line within the embedding space. A notable representative work in this category is MixGCF~\cite{Huang2021MixGCF}, which incorporates the \textit{mixup} technique~\cite{mixup}. MixGCF employs a mixing coefficient sampled from the beta distribution to blend the positive item and the sampled negative item. This blending process creates a linear interpolation between the positive and negative items, generating a pseudo-negative item located precisely on the line connecting the positive and negative instances. By acquiring a challenging negative item, the model gains improved discriminative capabilities between positive and negative items.
    \item This paper introduces the novel \textbf{Area-wise} sampling approach ($i_5$--$i_8$), which involves selecting a pseudo-negative item that resides within a specific area within the embedding space. Figure~\ref{fig:illustration} serves as an illustrative example. The hard negative area, depicted as a grey square, represents the region bounded by the dimensions of a positive item and a point-wise sampled negative item. When compared to Point-wise and Line-wise sampling methods, the Area-wise sampling technique offers a more extensive exploration space and greater flexibility. It allows for sampling from the hard negative area, which provides increased capacity for differentiating between positive and negative instances in multiple dimensions, thereby offering varying degrees of hardness.
\end{enumerate}

The proposed Area-wise sampling method faces several significant challenges that must be addressed. Firstly, the definition of boundaries for the hard negative area is crucial to the effectiveness of this approach. The sampling area plays a pivotal role, as an excessively large area may fail to provide informative negative items, while an overly small area could result in false negative item generation~\cite{ding2020simplify}.
The second challenge involves enabling Area-wise sampling within the determined hard negative area. Previous methods have primarily focused on sampling from discrete spaces or utilizing the \textit{mixup} technique to generate Line-wise pseudo-negative items. The ability to sample within a continuous area represents a novel research question in this context.
Lastly, given the flexibility introduced by sampling within an area, an additional challenge is effectively regularizing the sampling method to generate informative pseudo-negative items. Developing appropriate regularization techniques is necessary to ensure the quality and relevance of the generated negative samples.

% How our method designs, and how to deal with those drawback
This paper presents Dimension Independent Negative Sampling (\modelname) as a solution to facilitate Area-wise sampling in the context of collaborative filtering. \modelname comprises three distinct modules, each designed to address the aforementioned challenges in a targeted manner.
Firstly, the proposed \textbf{Hard Boundary Definition} module within \modelname determines the appropriate boundaries for the hard negative sampling area. It accomplishes this by selecting the item with the highest dot-product similarity to the positive item, thus establishing a boundary that closely aligns with the positive instance.
Following the establishment of boundaries, \modelname introduces a novel \textbf{Dimension Independent Mixup} module, enabling the sampling of items from the corresponding area. This module employs distinct linear interpolation weights for different dimensions, thereby extending the Line-wise sampling approach to support Area-wise sampling.
Lastly, \modelname proposes the \textbf{Multi-hop Pooling} module to regularize the sampling process and generate informative pseudo-negative items based on multiple hops of neighborhood information. By leveraging this module, \modelname achieves effective regularization, thereby enhancing the quality and relevance of the sampled negative instances.
By integrating these three modules, \modelname samples from the hard negative area successfully, and yield superior performance compared to other negative sampling methods when applied to various backbone models. The contributions of this work can be summarized as follows.
\begin{itemize}[leftmargin=*]
    \item Novel Perspective: We offer a fresh perspective from the embedding space to comprehend and analyze existing negative sampling methods, providing insights into their mechanisms.
    \item Area-wise Sampling: We are the first to introduce Area-wise negative sampling, recognizing the challenges, and presenting corresponding solutions.
    \item \modelname: We propose \modelname as the first Area-wise negative sampling method, enabling highly flexible sampling over an area. This novel approach surpasses existing methods and achieves state-of-the-art performance in collaborative filtering tasks.
    \item Experimental Validation: We conduct extensive experiments on three real-world datasets, employing different backbone models. The results demonstrate the effectiveness and superiority of \modelname, confirming its performance enhancement compared to other negative sampling methods.
\end{itemize}

The remaining paper is organized as follows. Section~\ref{sec:prelim} represents the preliminaries of this paper. Section~\ref{sec:method} illustrates different parts of \modelname in detail. Section~\ref{sec:experiment} conducts experiments comparing other methods and further experiments to show the effectiveness of \modelname. Section~\ref{sec:related_work} represents the most related works for reference, and we conclude \modelname and discuss future research directions in Section~\ref{sec:conclusion}.

\begin{figure*}[h]
    \centering
    \includegraphics[width=.9\textwidth]{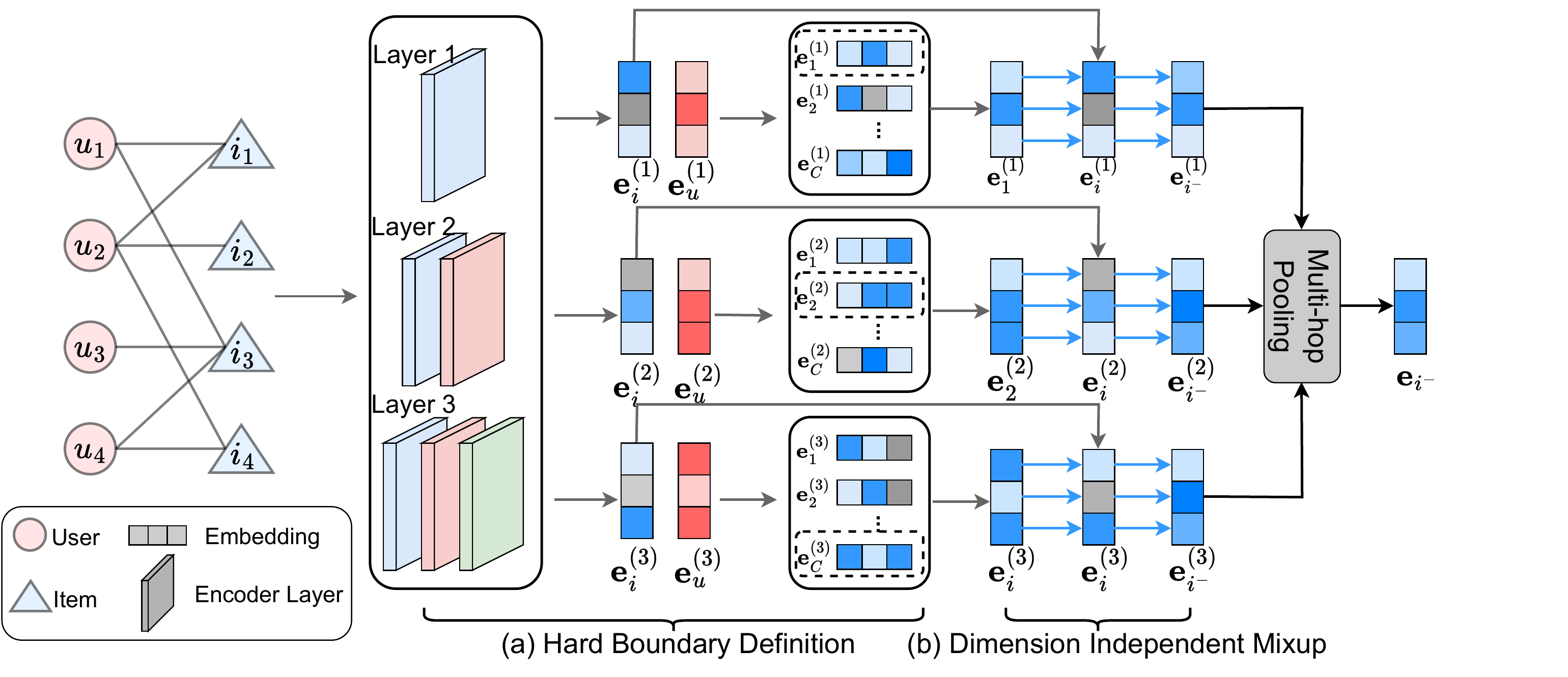}
    \caption{Main framework of \modelname. For each observed interaction $(u,i)$, \modelname first encodes $u$'s and $i$'s information within the graph with different numbers of GNN layers. For the output of each layer, (a) \modelname defines the sampling area by the Hard Boundary Definition module and then (b) mixes the sampled item with the corresponding positive item via the proposed Dimension Independent Mixup module. Finally, (c) \modelname generates the synthetic negative item by integrating the negative signal from different hops of neighborhoods with a Multi-hop Pooling module.}
    \label{fig:main_framework}
\end{figure*}

\section{Preliminaries}\label{sec:prelim}
In this section, we first formulate the collaborative filtering (CF) problem, and then illustrate the negative sampling problem in CF-based model training.

\subsection{Collaborative Filtering}
Collaborative filtering aims to predict user's preferences based on users' historical interactions. It has been shown as an effective and powerful tool~\cite{mao2021simplex,chae2018cfgan} for RecSys. With implicit feedback, we have a set of users $\mathcal{U} = \{u_1,u_2,...,u_{\left | \mathcal{U} \right|}\}$, a set of items $\mathcal{I} = \{i_1,i_2,...,i_{\left | \mathcal{I} \right|}\}$, and the observed user-item interactions $\textbf{R} \in \mathbb{R^{\left| \mathcal{U} \right| \times \left| \mathcal{I} \right|}}$, where $R_{u,i}=1$ if user $u$ has interacted with item $i$, or $R_{u,i}=0$ otherwise. Learning from historical interactions, current advanced CF-based methods~\cite{Wang2019NGCF,He2020lightgcn} learn an encoder function $f(\cdot)$ to map each user/item into a dense vector embedding, \ie, $f(u),f(i) \in \mathbb{R}^d$, where $d$ is the dimension of vector embedding. The predicted score from $u$ to $i$ is then calculated as the similarity of two vectors (\eg, dot product similarity, $R_{u,i}=f(u)^{\top} f(i)$). Then these methods rank all items based on the prediction score and select the top $k$ items $\{i_1, i_2,..., i_k\}$ as recommendation results for each user.

\subsection{Graph Neural Network for Recommendation}
In real-life applications, users can only interact with a limited number of items, which leads to the data sparsity problem~\cite{zheng2019deep}. To alleviate the problem, the most advanced CF-based models~\cite{He2020lightgcn,wu2021self,zhang2023apegnn} explicitly utilize high-order connections by representing the historical interactions as a user-item bipartite graph $\mathcal{G}=(\mathcal{V}, \mathcal{E})$, where $\mathcal{V}= \mathcal{U}\cup \mathcal{I}$ and there is an edge $(u,i) \in \mathcal{E}$ between $u$ and $i$ if $R_{u,i}=1$. By utilizing graph neural network (GNN) as the encoder function $f(\cdot)$, these methods learn the representations of user/item embeddings by aggregating information from their neighbors, so that connected nodes in the graph structure tend to have similar embeddings.
The operation of a general GNN computation can be expressed as follows:
\begin{align}
    \mathbf{e}_{u}^{(l+1)} = \mathbf{e}_{u}^{(l)} \oplus \text{AGG}^{(l+1)} (\{ \mathbf{e}_{i}^{(l)} \mid i \in \mathcal{N}_u \}),
\end{align}
where $\mathbf{e}_{u}^{(l)} \in \mathbb{R}^{D}$ is $u$'s embedding on the $l$-th layer, $\mathcal{N}_u$ is the neighborhoods of $u$, $\mbox{AGG}^{(l)}(\cdot)$ is a function that aggregates neighbors' embeddings into a single vector for layer $l$, and $\oplus$ combines $u$'s embedding with its neighbor's information. AGG$(\cdot)$ and $\oplus$ can be simple or complicated functions. Item is calculated in the same way.

After stacking $L$ layer GNN convolution, we can obtain $L$ user/item embedding from different GNN layers. Take the user as an example; we will obtain $(\mathbf{e}_{u}^{(1)}, \mathbf{e}_{u}^{(2)},..., \mathbf{e}_{u}^{(L)})$ for each user after the aggregation. $\mathbf{e}_{u}^{(l)}$ encodes the information within $l$-hop neighborhoods, which provides a unique user representation with local influence scope. A pooling function (\eg, attention, mean, sum) is used to combine them together $\mathbf{e}_{u}=\text{Pool}(\mathbf{e}_{u}^{(1)}, \mathbf{e}_{u}^{(2)},..., \mathbf{e}_{u}^{(L)})$.

\subsection{The Negative Sampling Problem}
The observed implicit feedback in RecSys only indicates the user's positive interest. Training with only positive labels would cause the model degradation without the ability to distinguish different items. Thus, the training of CF-based models involves the negative sampling procedure to provide the negative signal. The corresponding training method trains the model to give higher scores to observed interactions while lower scores to negative sampled interactions. The most renowned is pair-wise BPR loss~\cite{Rendle2012BPR}:
\begin{equation}\label{eq:bpr}
    \mathcal{L}_{BPR}=\frac{1}{|\mathcal{E}|}\sum_{(u,i)\in \mathcal{E}}-\text{log}(\text{sigmoid}(s(u,i)-s(u,i^{-}))),
\end{equation}
where $s(u,i)$ is the predicted rating score from $u$ to $i$, and $i^{-}$ is the negative sampled item. The quality of sampled $i^{-}$ critically impacts the model performance. Hard negative samples can effectively assist the model in learning better boundaries between positive and negative items~\cite{pinsage}. 
In terms of sampling methods, discrete sampling\cite{Rendle2012BPR,DNS,Wang2017IRGAN:,pinsage} can be considered as point-wise sampling, which selects negative items from a discrete sample space. Negative sampling based on mixup\cite{Huang2021MixGCF,SoftBPR}, on the other hand, can be classified as line-wise sampling because this method performs a linear interpolation between two items to obtain a new negative sample along the diagonal line. Both point-wise sampling and line-wise sampling, in fact, do not adequately approximate the positives in the sample space and, therefore, cannot train a more powerful encoder. In this paper, we propose the first area-wise sampling method that greatly extends the sampling area.

\section{Proposed Method}\label{sec:method}
% 这里可以借鉴Soft BPR Loss for Dynamic Hard Negative Sampling in Recommender Systems这篇文章里的第三章写法，它从最原始的mixup开始讲起，一直讲到最近Mixup的最新应用。
% 良伟师兄原话
This section illustrates the proposed \modelname that enables area-wise negative sampling. As Figure~\ref{fig:main_framework} demonstrates, \modelname mainly consists of three parts, \textit{i.e.}, Hard Boundary Definition, Dimension Independent Mixup, and Multi-hop Pooling.

% 下面是修改意见后 (comment: 章节开始不需要讲着么细）
% \textcolor{red}{
% Based on the analysis of MCNS \cite{MCNS} theory, the distribution of negatives should be closer to positives the better, and several recent studies\cite{Huang2021MixGCF,SoftBPR} have made similar findings. While these existing methods have shown improvements in model performance, we argue that they do not fully explore the vicinity of positive items. Therefore, we propose DINS, which uses area-sampling to enable negative samples to approximate positive samples both locally and overall in the item space. The structure of \modelname is illustrated in Figure \ref{fig:main_framework}. The \modelname mainly consists of three parts, \textit{i.e.}, Hard Boundary Definition, Dimension Independent Mixup, and Multi-hop Pooling.}

% In the Hard Boundary Definition module, DINS first needs to determine a negative item to form a suitable negative sampling area with the positive item. Then DINS uses the Dimension Independent Mixup module to bring the negative item closer to the positives, both locally and globally. Finally DINS can extend the method of area negative sampling to GNN-based encoders with the Hard Boundary Definition module.

\subsection{Hard Boundary Definition}\label{sec:boundary}
The first question to answer in the proposed area-wise negative sampling is how to define the continuous sampling area. The hard boundary definition module is shown in Figure~\ref{fig:main_framework}(a). It samples a boundary item to define the sampling area. For each interaction $(u,i)$, \modelname defines the area by sampling a boundary item $i_{*}$. In each sampling, \modelname pre-samples a candidate set $\mathcal{C}=(i_{1},i_{2}, ..., i_{|\mathcal{C}|})$ to reduce the computation workload as previous researches~\cite{ding2020simplify,DNS}. Then \modelname selects the item with the highest dot-product with $u$ to define the boundary:
\begin{equation}\label{equation:HNS}
    i_{*} = \argmax_{i}(\mathbf{e}_{u}^{\top}\mathbf{e}_{i_1}, \mathbf{e}_{u}^{\top}\mathbf{e}_{i_2},...,\mathbf{e}_{u}^{\top}\mathbf{e}_{i_{|\mathcal{C}|}}).
\end{equation}
The boundary item $i_{*}$ defines the hard negative area together with the corresponding positive item $i$ in the embedding space. As shown in Figure~\ref{fig:illustration}, the hard negative area is defined as the space between $i$ and $i_{*}$. The majority of previous research~\cite{DNS,ding2020simplify,Rendle2012BPR} samples discrete existing items. While MixGCF~\cite{Huang2021MixGCF} allows the sampling on continuous space along the diagonal line from $i$ to $i_{*}$, it still leaves the large-volume hard negative area un-explored.

\subsection{Dimension Independent Mixup}\label{sec:mixup}
\begin{figure}[h]
    \centering
    \includegraphics[width=.3\textwidth]{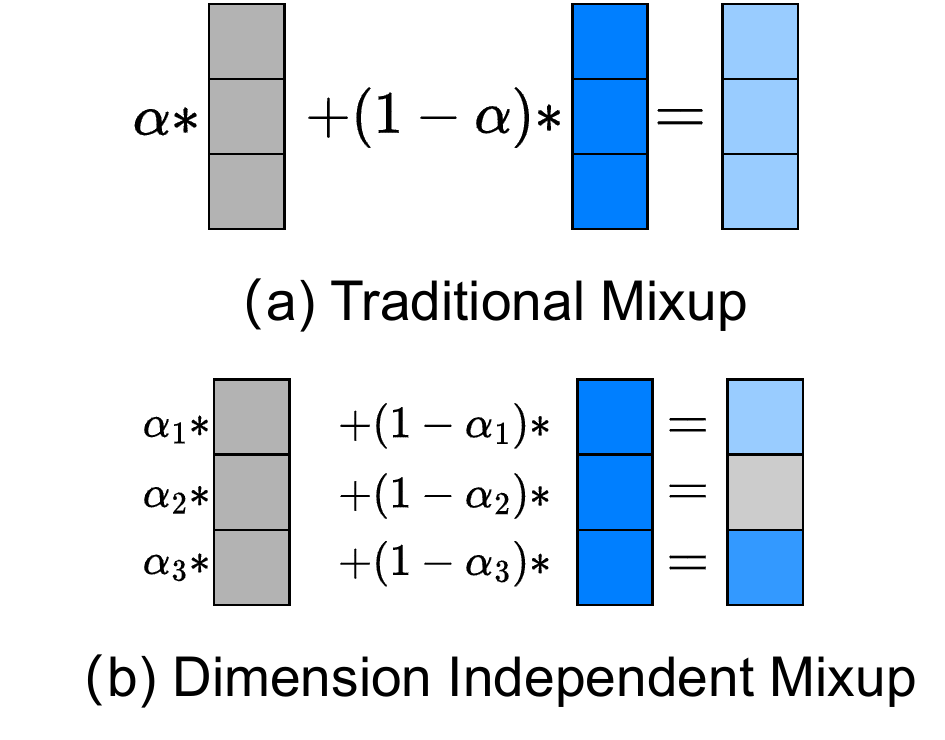}
    \caption{(a) Traditional Mixup assigns the same interpolation weights on all dimensions. (b) The proposed Dimension Independent Mixup assigns different interpolation weights on different dimensions.}
    \label{fig:mixup}
\end{figure}
The second question to answer is how to explore the non-diagonal space within the hard negative area and extend the line-wise to area-wise sampling. To achieve this, we propose the novel Dimension Independent Mixup method in Figure~\ref{fig:main_framework}(b). It enables a dimension independent mix between the boundary item $i_{*}$ and positive item $i$. 

A further comparison to the traditional Mixup is shown in Figure~\ref{fig:mixup}. Line-wise sampling methods~\cite{Huang2021MixGCF,SoftBPR} utilize the traditional Mixup to generate a negative item. As shown in Figure~\ref{fig:mixup}(a), traditional Mixup synthesizes the negative item via linear interpolation of the positive and boundary items with a unified weight $\alpha$ on all dimensions. With all dimensions having the same linear interpolation, it can only generate a synthetic item falling on the line between the two mixed items in the continuous embedding space. 
Figure~\ref{fig:mixup}(b) shows the proposed Dimension Independent Mixup, which is the core idea to support area-wise negative samples. It mixes the two items dimension independently by calculating specific interpolation weights for different dimensions. It is more flexible and increases the exploration space from a line to the whole hard negative area.

For each interaction $(u,i)$, Section~\ref{sec:boundary} samples a boundary item $i_{*}$ to define the sampling area. The mixup for the $d$-th dimension is calculated as:
\begin{equation}\label{equation:mix}
    e_{i^{-}}^d = \alpha_d*e_{i_{*}}^{d} + (1-\alpha_d)*e_{i}^d,
\end{equation}
where $e_i^d$ is the $d$-th dimension value of $\mathbf{e}_i$. $\alpha_d$ is the interpolation weight for the $d$-th dimension, which is calculated as :
\begin{equation}\label{equation:Wn}
    \alpha_d = \frac{\exp(e_u^d * e_{i_{*}}^d)}{\exp(e_u^d * e_{i_{*}}^d) + \beta*\exp(e_u^d * e_i^d)},
\end{equation}
where $e_u^d$ is the $d$-th dimension value of $\mathbf{e}_u$, and $\beta>0$ is a hyper-parameter to tune the relative weight for the mixup. A larger $\beta$ leads to a smaller $\alpha_d$, and $e_{i^{-}}^d$ is more similar with positive item.
We concatenate the dimensions together to obtain the mixed item:
\begin{equation}\label{equation:concat}
    \mathbf{e}_{i^{-}} = \text{Concat}(e_{i^{-}}^1, e_{i^{-}}^2,...,e_{i^{-}}^D)
\end{equation}
$\mathbf{e}_{i^{-}}$ is the generated negative item embedding for the interaction $(u,i)$. It can be directly used to calculate the ranking loss in Equation~\ref{eq:bpr}.
Calculating the mixup weight independently for each dimension extends the exploration area from a line to the whole hard negative area. It greatly improves the final recommendation performance by providing more varied items, as illustrated in Section~\ref{sec:experiment}.

\subsection{Multi-hop Pooling}
We further extend \modelname to better support graph neural network (GNN) based collaborative filtering methods~\cite{He2020lightgcn,yang2021consisrec}. These methods are shown to be effective by explicitly considering the high-order connections and encode user/item high-order neighborhood information into dense vectors. The output of different GNN layers encodes neighborhoods within different hops. Regarding the negative sample, MixGCF~\cite{Huang2021MixGCF} also proves to be effective when considering different hops of neighborhoods. To utilize the high-order information, \modelname samples negative items based on each dense vector extracted by different GNN layers and utilizes a pooling module to obtain the final negative item embedding.

When we utilize GNN as the encoder model $f(\cdot)$, we can obtain one dense embedding from each GNN layer to encode the corresponding neighborhoods' information. After $L$ layers GNN encoding on the user-item bipartite graph $\mathcal{G}$, we obtain $L$ dense representations for each user $u$ and item $i$:
\begin{equation}
\begin{split}
    f(u) &= (\mathbf{e}_{u}^{(1)}, \mathbf{e}_{u}^{(2)},..., \mathbf{e}_{u}^{(L)}) \\
    f(i) &= (\mathbf{e}_{i}^{(1)}, \mathbf{e}_{i}^{(2)},..., \mathbf{e}_{i}^{(L)}).
\end{split}
\end{equation}
Following Section~\ref{sec:boundary} and Section~\ref{sec:mixup}, we synthesize a negative item embedding $\mathbf{e}_{i^{-}}^{(l)}$ based on $\mathbf{e}_{u}^{(l)}$ and $\mathbf{e}_{i^{-}}^{(l)}$. Then we obtain the final negative sample via a Pooling function:
\begin{equation}\label{eq:pool}
    \mathbf{e}_{i^{-}}=\text{Pool}(\mathbf{e}_{i^{-}}^{(1)}, \mathbf{e}_{i^{-}}^{(2)},..., \mathbf{e}_{i^{-}}^{(L)}),
\end{equation}
where the Pool can be any function pools multiple tensors into a single tensor. For simplicity, we test mean pooling and concatenation functon for all datasets.

\subsection{Complexity Analysis}
The training process of \modelname is illustrated step by step in Algorithm~\ref{algo}. The time complexity for each sampling process comes from the three proposed modules. Hard Boundary Definition module takes $O(MD)$, where $M$ is the candidate set budget and $D$ is the dimension size. Dimension Independent Mixup module takes $O(D)$. Thus, the time complexity without considering multi-hop neighbors is $O(MD)$. When considering the multi-hop neighbors, \modelname needs an extra sampling procedure for each GNN layer. The total time complexity is then $O(LMD)$, where $L$ is the number of GNN layers. To be noted that, the time complexity is the same as MixGCF, but \modelname extends to a much larger exploration space.

\renewcommand{\thealgorithm}{1} % 根据需要设定算法编号
\begin{algorithm}
  \caption{The training process with \modelname}
  \label{algo}
  \begin{algorithmic}[1]
  \REQUIRE Training set$\mathcal{G}$, Recommendation user/item encoder $f(\cdot)$, a candidate set budget $M$, a parameter $\beta$ to control positive mixing
  \FOR{$t= 1,2,3, \dots ,$ to $T$}
  \STATE Sample a mini-batch of user-item positive pairs$\{(u,i)\}$.
  \STATE Initialize loss $\mathcal{L}_{BPR}=0$.
  \FOR{each $(u,i)$ pair}
  \STATE Get the user and positive item embeddings by encoder $f(\cdot)$.
  \STATE Get ID embedding of uniformly sampled $M$ negatives as the candidate set.
  \STATE Get the boundary negative item $i_{*}$ by (\ref{equation:HNS}).
  \STATE Calculate the independent dimensional weight matrix of boundary item $i_{*}$ by (\ref{equation:Wn}).
  \STATE Synthesize a hard negative item $e_{i^{-}}$ by (\ref{equation:mix}) and (\ref{equation:concat}).
  \IF{$f(\cdot)$ is GNN-based model}
    \STATE Further synthesize hard negative item for output of different GNN layers
    \STATE Pooling the synthesized negative items by (\ref{eq:pool})
  \ENDIF
  \STATE Calculate $\mathcal{L}_{BPR}$ by (\ref{eq:pool}).
  \ENDFOR
  \STATE Update $\theta$ by descending the gradients $\bigtriangledown _{\theta}\mathcal{L}_{BPR}$ .
  
  \ENDFOR
    % 伪代码算法的内容
  \end{algorithmic}
\end{algorithm}

\section{Experiments}\label{sec:experiment}
This section empirically evaluates the proposed \modelname on three real-world datasets with three different backbones. The goal is to answer the four following research questions (RQs). 
\begin{itemize}[leftmargin=*]
    \item \textbf{RQ1:} Can \modelname provide informative negative samples to improve the performance of recommendation? 
    \item \textbf{RQ2:} Does every module contributes to the effectiveness?
    \item \textbf{RQ3:} What is the impact of different hyper-parameters on \modelname?
    \item \textbf{RQ4:} Is \modelname really supporting area-wise negative sampling? 
\end{itemize}

\subsection{Experimental Setup}

\subsubsection{Datasets} 
\begin{table}[]
\caption{Statistics of the datasets.}
\begin{tabular}{c|cccc}
\toprule
Dataset  & \#User & \#Items & \#Interactions & Density \\ \hline 
Alibaba  & 106,042 & 53,591   & 907,407   & 0.016\% \\ 
Yelp2018 & 31,668  & 38,048   & 1,561,406    & 0.13\%  \\ 
Amazon   & 192,403 & 63,001   & 1,689,188    & 0.014\% \\ 
\bottomrule
\end{tabular}
\label{table1}
\end{table}

% 为了评估NGCF的有效性，我们在三个基准数据集： Gowalla、Yelp2018和Amazon-book上进行了实验。
% 这些数据集是公开的，在领域、大小和稀疏度方面都有所不同。我们在表1中总结了三个数据集的统计数据。
For a fair comparison, we also evaluate \modelname on three benchmark datasets: Alibaba~\cite{MCNS,Huang2021MixGCF}, Yelp2018~\cite{Wang2019NGCF,Wang2019NGCF}, and Amazon~\cite{MCNS,Huang2021MixGCF} following previous research~\cite{Huang2021MixGCF}. We also follow the same training, validation, and testing split setting~\cite{Wang2019NGCF,MCNS,Huang2021MixGCF}.
The detailed Statistics of three public datasets are shown in Table~\ref{table1}, which exhibits the variation in scale and sparsity. %Yelp2018 demonstrates the highest level of density among these three datasets under consideration, while the other two datasets exhibit relatively lower levels of density.

\subsubsection{Baselines}
To validate the effectiveness of DINS, we chose three backbone networks, LightGCN~\cite{He2020lightgcn} and NGCF~\cite{Wang2019NGCF} as GNN-based encoders and MF~\cite{Rendle2012BPR} as the non-GNN-based encoder. Additionally, we selected six negative sampling methods for comparison.
\begin{itemize}[leftmargin=*]
    \item \textbf{Popularity}: It samples negative items by assigning a higher sampling probability of more popular items.
    \item \textbf{RNS}~\cite{Rendle2012BPR}: Random negative sampling (RNS) strategy is a widely used approach, which applies a uniform distribution to sample an item that the user has never interacted with.
    \item \textbf{DNS}~\cite{DNS}: Dynamic negative sampling (DNS) strategy adaptively selects the highest-scoring negative item by the current recommendation model among randomly selected items. Such a negative item is considered a hard negative item for training.
    \item \textbf{IRGAN}~\cite{Wang2017IRGAN:}: It is a GAN-based strategy for generating negative sampling distribution.
    %\item \textbf{MixGCF}~\cite{Huang2021MixGCF}: MixGCF is the state-of-the-art sampling method based on Mixup, which applies positive mixing and hop mixing to synthesize new negative items.
    \item \textbf{MixGCF}~\cite{Huang2021MixGCF}: MixGCF is a graph-based negative sampling method, which applies positive mixing and hop mixing to synthesize new negative items. However, MF is not a GNN-based encoder, so we only use positive mixing under MF. We mark this MixGCF which uses hop mixing only with * in table\ref{performance comprision}.
    \item \textbf{DENS}~\cite{DENS}: Disentangled negative sampling (DENS) effectively extracts relevant and irrelevant factors of items, later employing a factor-aware strategy to select optimal negative samples.
\end{itemize}

\subsubsection{Evaluation Method} 
% We use the two most commonly used evaluation metrics in recommender systems to measure the effectiveness of our model.
We use  Recall@K and NDCG@K to evaluate the performance of the top-K recommendation of our model, both of which are widely used in the recommendation system. By default, the value of K is set as 20.
We present the average metrics for all users in the test set, calculating these metrics based on the rankings of non-interacted items.
Following prior research~\cite{Wang2019NGCF, He2020lightgcn}, we employ the complete full-ranking technique, which involves ranking all items that have not yet been interacted with by the user. 

\subsubsection{Experimental Settings} 
% We implement our \modelname and other baseline models based on Pytorch 2.0 and Python 3.8. Both the user and item embedding sizes are fixed to 64 for backbone recommendation models (NGCF, LightGCN, and MF). To achieve better performance, we use Xavier\cite{Glorot2010Xavier} to initialize embedding parameters for all encoders, and the number of aggregation layers in NGCF or LightGCN is set as 3 by default. Adam~\cite{Kingma2014Adam:} is used to optimize all encoders. We set the batch size as 2048 for LightGCN and MF and 1024 for NGCF. For the remaining hyper-parameters, we used the grid search technique to find the optimal settings for each recommender: the learning rate is searched in \{0.0001, 0.001, 0.01\}, coefficient of weight decay is tuned in \{$1e^{-4}, 1e^{-5}, 1e^{-6}$\} . Moreover, the size of the candidate pool for DNS, MixGCF, and our \modelname is searched in \{8,16,32,64\}, and the hyper-parameter $\beta$, which is used in our method is searched from 0 to 10. 
We implement our \modelname and other baseline models based on Pytorch, and the embedding size is fixed to 64. To achieve better performance, we use Xavier\cite{Glorot2010Xavier} to initialize embedding parameters for all encoders, and the number of aggregation layers in NGCF or LightGCN is set as 3. Adam~\cite{Kingma2014Adam:} is used to optimize all encoders. We set the batch size as 2048 for LightGCN and MF and 1024 for NGCF. For the remaining hyper-parameters, we used the grid search technique to find the optimal settings for each recommender: the learning rate is searched in \{0.0001, 0.001, 0.01\}, coefficient of weight decay is tuned in \{$1e^{-4}, 1e^{-5}, 1e^{-6}$\} . Moreover, the size of the candidate pool for DNS, MixGCF, and \modelname is searched in \{8,16,32,64\}, and the hyper-parameter $\beta$, which is used in our method is searched from 0 to 10.

\subsection{RQ1: Performance Evaluation}
\begin{table*}[]
\caption{Performance Comparision. The best and runner-ups are marked in bold and underlined separately.}
\label{performance comprision}
% \usepackage{multirow}
% \usepackage[normalem]{ulem}

\begin{tabular}{c|c|cc|cc|cc}
\hline
{\multirowcell{2}{Backbone \\ Model}} & {\multirowcell{2}{Sampling \\ Method}} & \multicolumn{2}{c}{Amazon}                                                 & \multicolumn{2}{c}{Alibaba}                                                    & \multicolumn{2}{c}{Yelp208}                                \\ \cline{3-4} \cline{5-6} \cline{7-8} 
 & {}                  & Recall@20       &  {NDCG@20}         & Recall@20      &  {NDCG@20}         & Recall@20     & NDCG@20              \\ \hline
\multirow{6}{*}{LightGCN} & {Popularity}  & 0.0323        &  {0.0153}          &  {0.0481}          &  {0.0231}          &  {0.0469}         & 0.0369               \\ 
 & {RNS}               & 0.0399          &  {0.0178}          &  {0.0550}          &  {0.0251}          &  {0.0605}         & 0.0493               \\ 
 & {DNS}               & 0.0453          &  {0.0211}          &  {0.0576}          &  {0.0258}          &  {{\ul 0.0706}}   & {\ul 0.0581}         \\ 
 & {IRGAN}             & 0.0338          &  {0.0150}          &  {0.0551}          &  {0.0255}          &  {0.0535}         & 0.0251               \\ 
 & {DENS}             &   0.0429        &  {0.0195}          &  {0.0637}          &  {0.0294}          &  {0.0560}         & 0.0457                \\ 
 & {MixGCF}            & {\ul 0.0456}    &  {{\ul 0.0214}}    &  {{\ul 0.0689}}    &  {{\ul 0.0332}}    &  {0.0691}         & 0.0565               \\ 
 & {\modelname}              & {\textbf{0.050}}    &  {\textbf{0.0236}} &  {\textbf{0.0764}} &  {\textbf{0.0358}} &  {\textbf{0.0738}} & \textbf{0.0604}      \\ \hline
 & Improvement         & 9.6\%                        & 10.3\%                      & 10.9\%                      & 7.9\%                       & 4.5\%                      & 4.0\%   \\ \bottomrule \bottomrule
\multirow{6}{*}{NGCF} & {Popularity}  &  {0.0115}          &  {0.0047}          &  {0.0180}          &  {0.0080}          &  {0.0253}         & 0.0196               \\ 
 &{RNS}               &  {0.0288}          &  {0.0119}          &  {0.0337}          &  {0.0144}          &  {0.0561}         & 0.0457               \\ 
 &{DNS}               &  {0.0304}          &  {0.0131}          &  {0.0475}          &  {0.0228}          &  {0.0634}         & 0.0520               \\ 
 &{IRGAN}             &  {0.0194}          &  {0.0078}          &  {0.0280}          &  {0.0116}          &  {0.0438}         & 0.0353               \\ 
 & {DENS}             &   0.0337        &  {0.0149}          &  {0.0383}          &  {0.0164}          &  {0.053}         &  0.0433              \\ 
 &{MixGCF}            &  {{\ul 0.0350}}    &  {{\ul 0.0150}}    &  {{\ul 0.0562}}    &  {{\ul 0.0268}}    &  {{\ul 0.0686}}   & {\ul 0.0567}         \\ 
 &{\modelname}         &  {\textbf{0.0379}} &  {\textbf{0.0163}}  &  {\textbf{0.0607}}  &  {\textbf{0.0277}}  &  {\textbf{0.0709}} & \textbf{0.0586}       \\ \hline
 & Improvement   & 8.3\%                      & 8.7\%                       & 8.0\%                       & 3.4\%                       & 3.4\%                      & 3.4\% \\ \bottomrule \bottomrule
\multirow{6}{*}{MF} & {Popularity}  &  {0.0148}          &  {0.0122}          &  {0.0215}          &  {0.0103}          &  {0.0382}         & 0.0317               \\ 
 &{RNS}               &  {0.0245}          &  {0.0104}          &  {0.0301}          &  {0.0144}          &  {0.0558}         & 0.0449               \\ 
 &{DNS}               &  {0.0320}          &  {0.0154}          &  {{\ul 0.0487}}    &  {{\ul 0.0240}}    &  {{\ul 0.0663}}   & {\ul 0.0547}         \\ 
 &{IRGAN}             &  {0.0281}          &  {0.0119}          &  {0.0307}          &  {0.0139}          &  {0.0412}         & 0.0338               \\ 
 & {DENS}             &    {0.0328}       &  {0.0147}          &  {0.0339}          &  {0.0163}          &  {0.0527}         &    {0.0431}            \\ 
 &{MixGCF*}            &  {\ul{0.0342}}                &  {{\ul 0.0156}}                &  {0.0480}          &  {0.0232}          &  {0.0642}         & 0.0525               \\ 
 &{\modelname}         &  {\textbf{0.0423}}          &  {\textbf{0.0197}}          &  {\textbf{0.0663}} &  {\textbf{0.0320}} &  {\textbf{0.0699}} & \textbf{0.0579} \\ \hline
 & Improvement   &  23.7\%                        &  26.3\%                        &  36.1\%               &  33.3\%               & 8.9\%                 & 10.3\% \\ \bottomrule
\end{tabular}

\end{table*}

We report the overall performance of the six baselines on the three backbones in Table \ref{performance comprision}. We can have the following observations:
\begin{itemize}[leftmargin=*]   
    % \item \modelname consistently outperforms all baselines by a large margin on three datasets across three classical encoders. Specifically, \modelname accomplishes remarkable improvements over the second-best baseline, especially on the Amazon and Alibaba datasets which improved Recall@20 by 23\%, and 36.1\%, respectively. This further demonstrates the effectiveness of area-wise sampling, as it enhances the performance of not only GNN-based encoders but also non-GNN-based encoders.
    \item \modelname outperforms all baselines by a large margin on three datasets across three encoders and accomplishes remarkable improvements over the second-best baseline, especially on the Amazon and Alibaba datasets which improved Recall@20 by 23\%, and 36.1\%, respectively. This further demonstrates the effectiveness of area-wise sampling, as it enhances the performance of not only GNN-based encoders but also non-GNN-based encoders.
    \item In most cases, the line-wise sampling method (MixGCF) is better than the rest point-wise sampling methods. It shows the advantage of extending exploration space from points to lines.
    % \item LightGCN consistently outperforms NGCF and MF in all cases, and NGCF performs better than MF in most cases, which demonstrates the effectiveness of using graph neural networks to encode multi-hop implicit collaborative filtering signals.
    \item \modelname exhibits greater improvement on the two datasets of lower density (Alibaba and Amazon), ranging from 8.0\% to 36.1\%. This highlights \modelname's ability to explore the continuous embedding space and effectively enhance performance on sparse datasets.
\end{itemize}

\begin{figure}
\centering
  \begin{subfigure}[b]{0.23\textwidth}
    \centering
    \includegraphics[width=\linewidth]{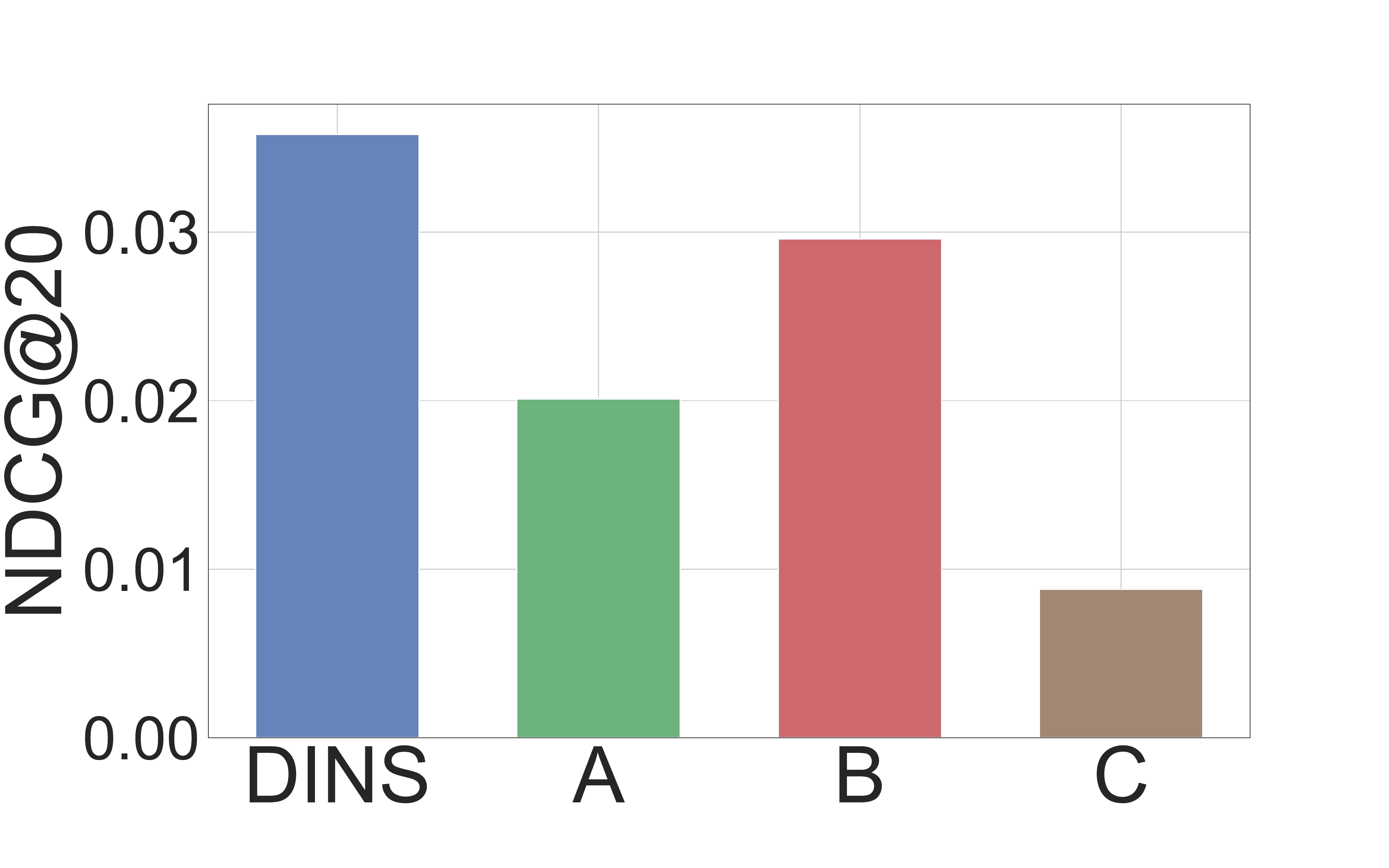}
    \subcaption{Alibaba-NDCG}
  \end{subfigure}
  \begin{subfigure}[b]{0.23\textwidth}
    \centering
    \includegraphics[width=\linewidth]{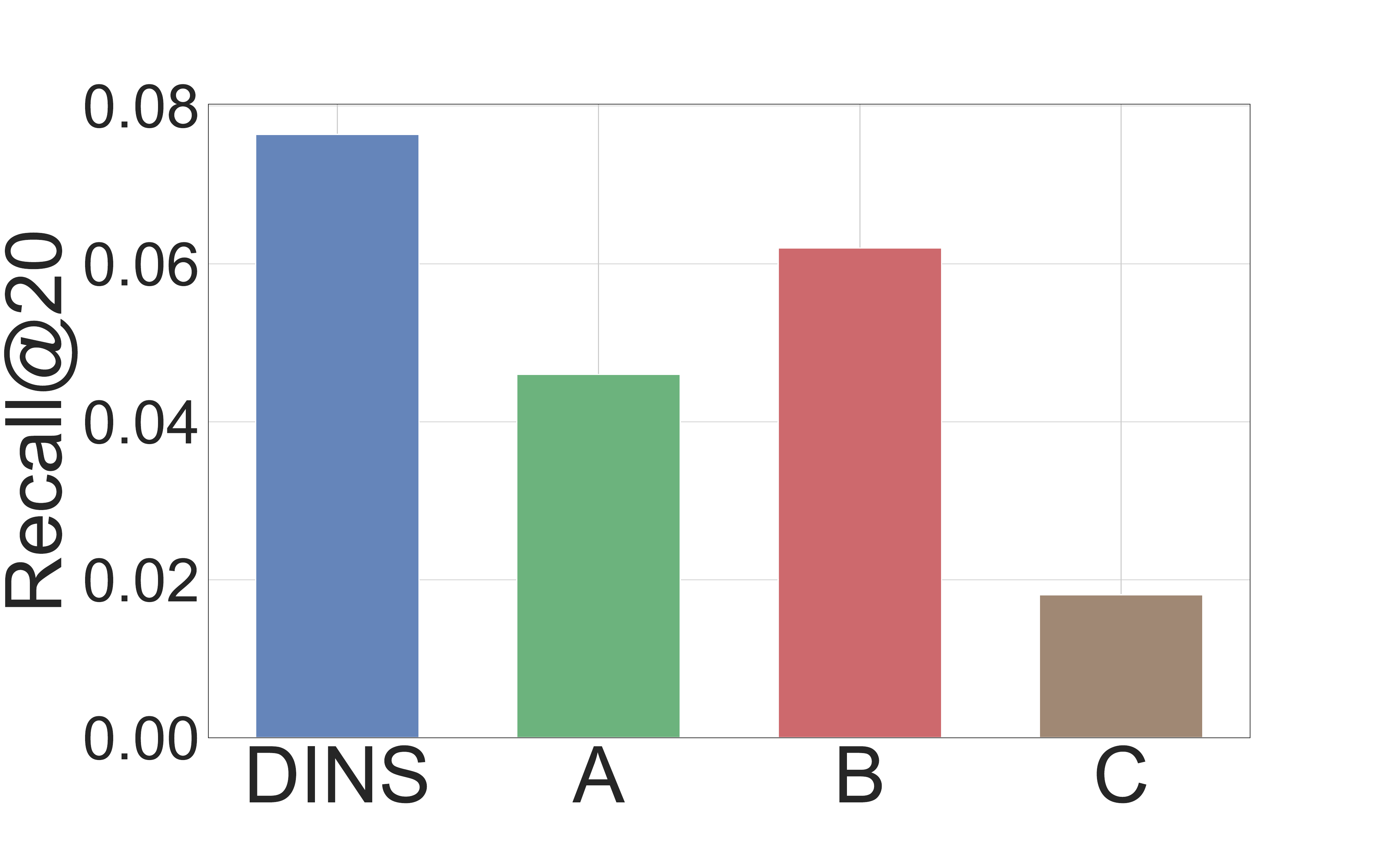}
    \subcaption{Alibaba-Recall}
  \end{subfigure}

  \begin{subfigure}[b]{0.23\textwidth}
    \centering
    \includegraphics[width=\linewidth]{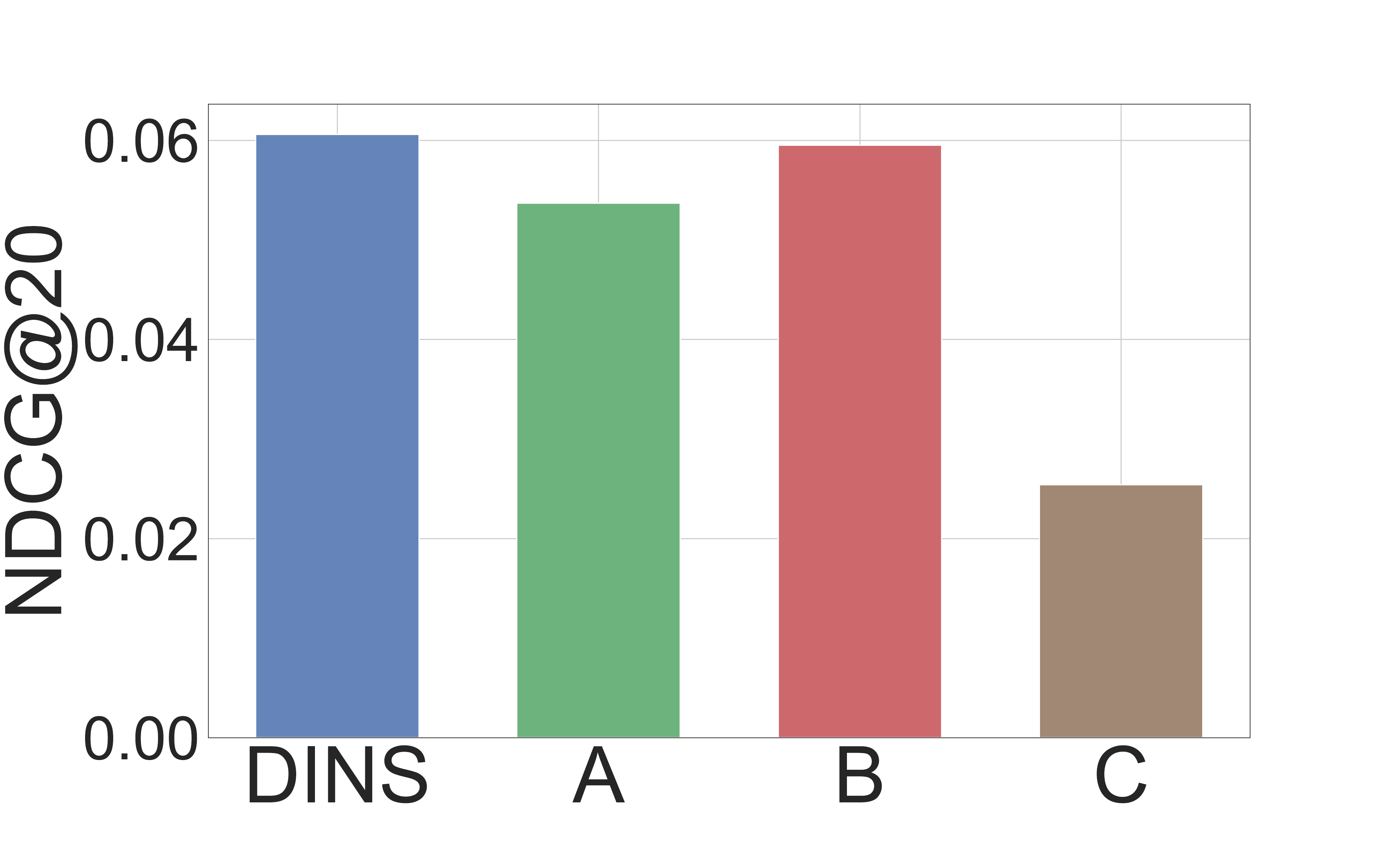}
    \subcaption{Yelp2018-NDCG}
  \end{subfigure}
  \begin{subfigure}[b]{0.23\textwidth}
    \centering
    \includegraphics[width=\linewidth]{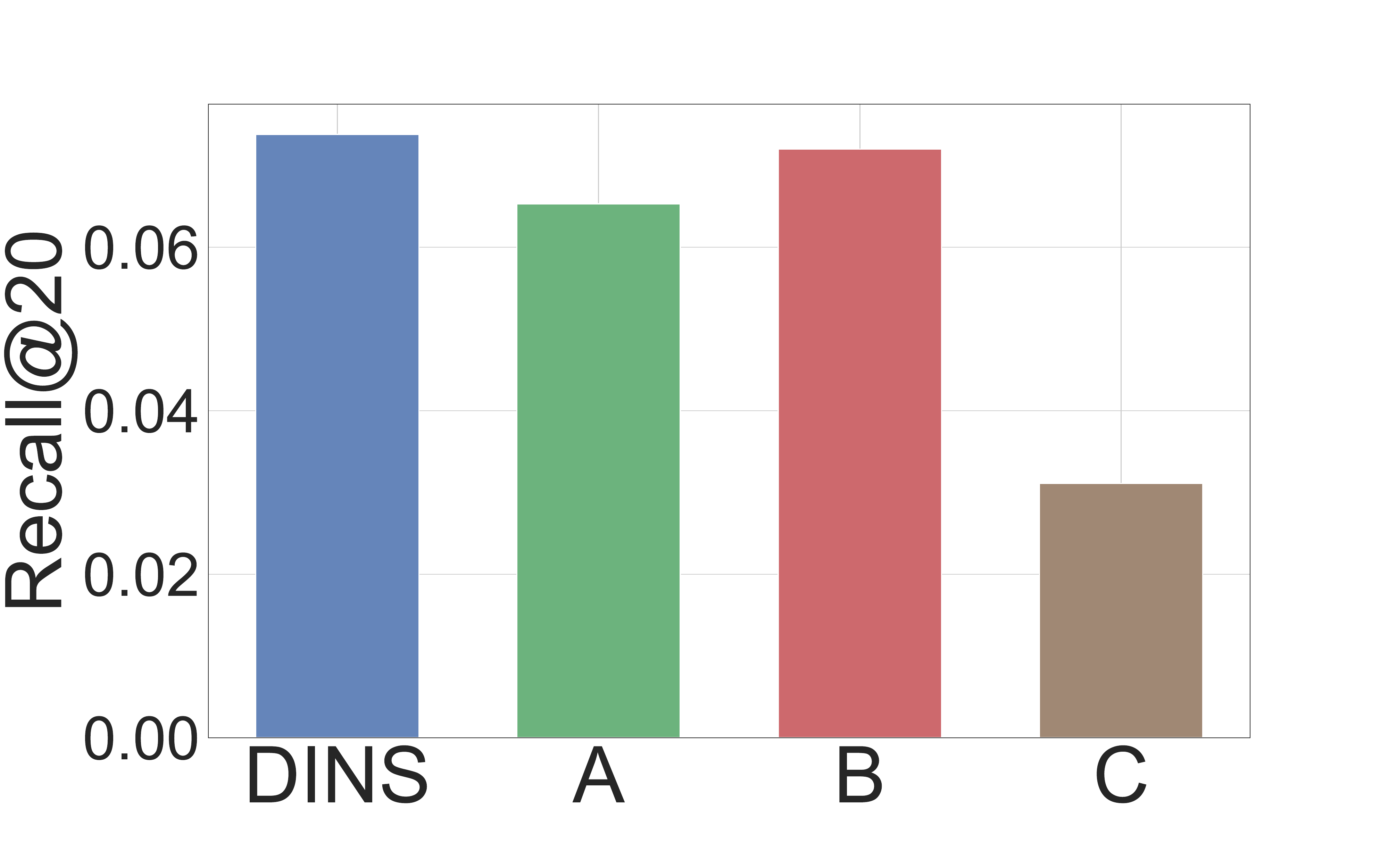}
    \subcaption{Yelp2018-Recall}
  \end{subfigure}
\caption{Ablation experiments on LightGCN.}
\label{fig:ablation}
\end{figure}

\begin{figure*}[t]
  \centering

% 第二个排版凡是方式
% 这里是第一排的位置，全部放Recall@20
  \begin{subfigure}[b]{0.23\textwidth}
    \centering
    \includegraphics[width=\linewidth]{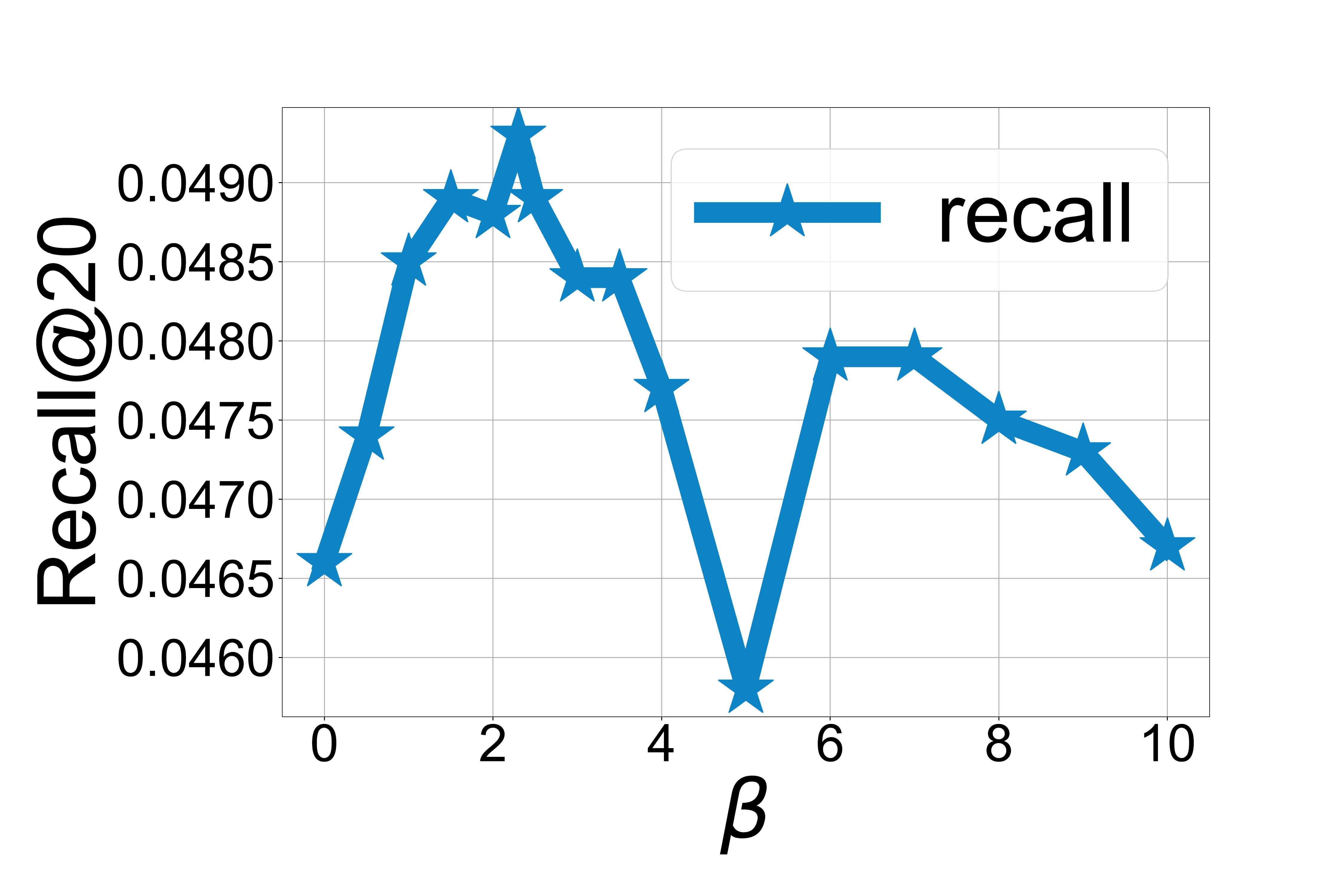}
    \caption{LightGCN-Amazon}
  \end{subfigure}
  \begin{subfigure}[b]{0.23\textwidth}
    \centering
    \includegraphics[width=\linewidth]{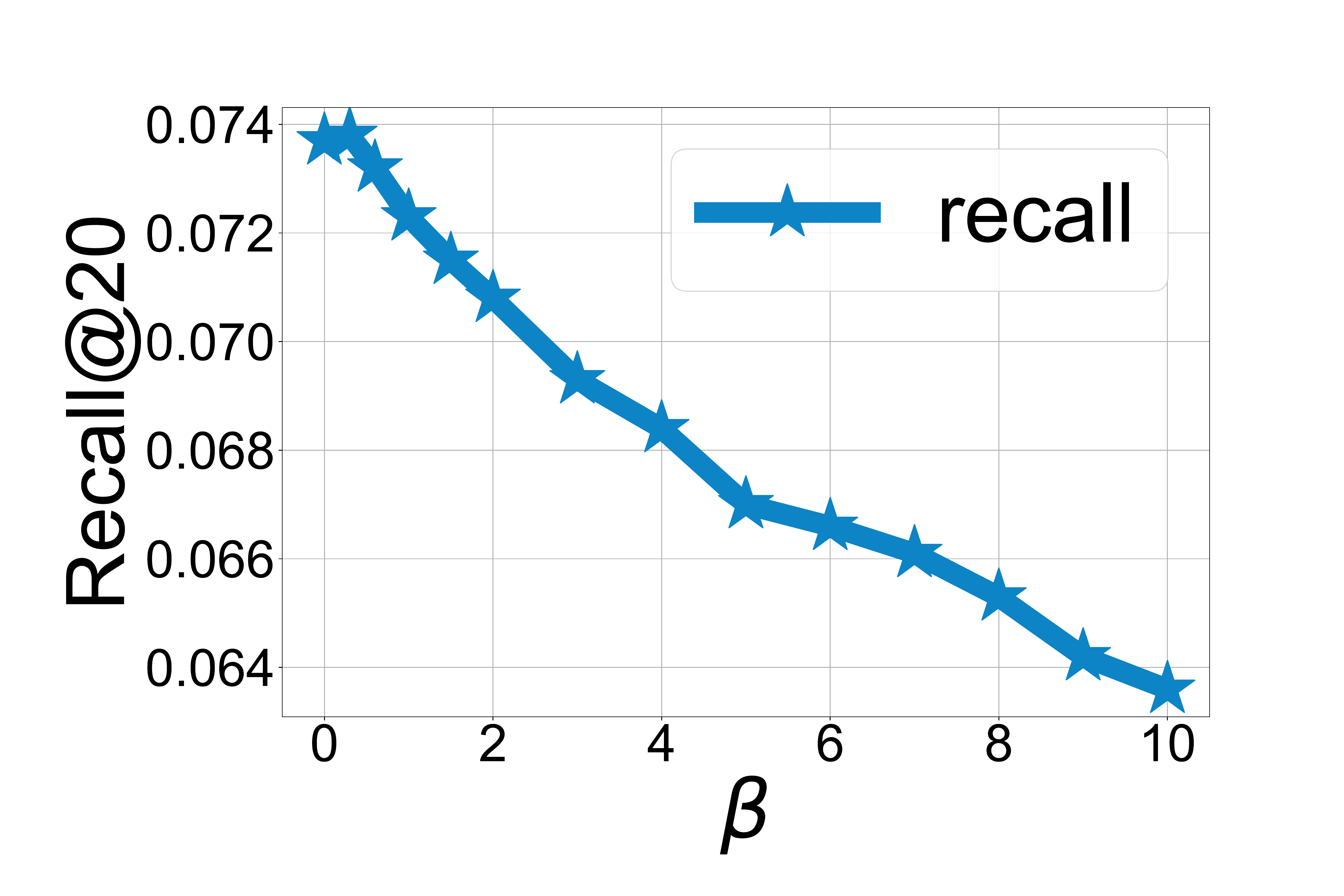}
    \caption{LightGCN-Yelp2018}
  \end{subfigure}
  % \caption{MF-Amazon}
  \begin{subfigure}[b]{0.23\textwidth}
    \centering
    \includegraphics[width=\linewidth]{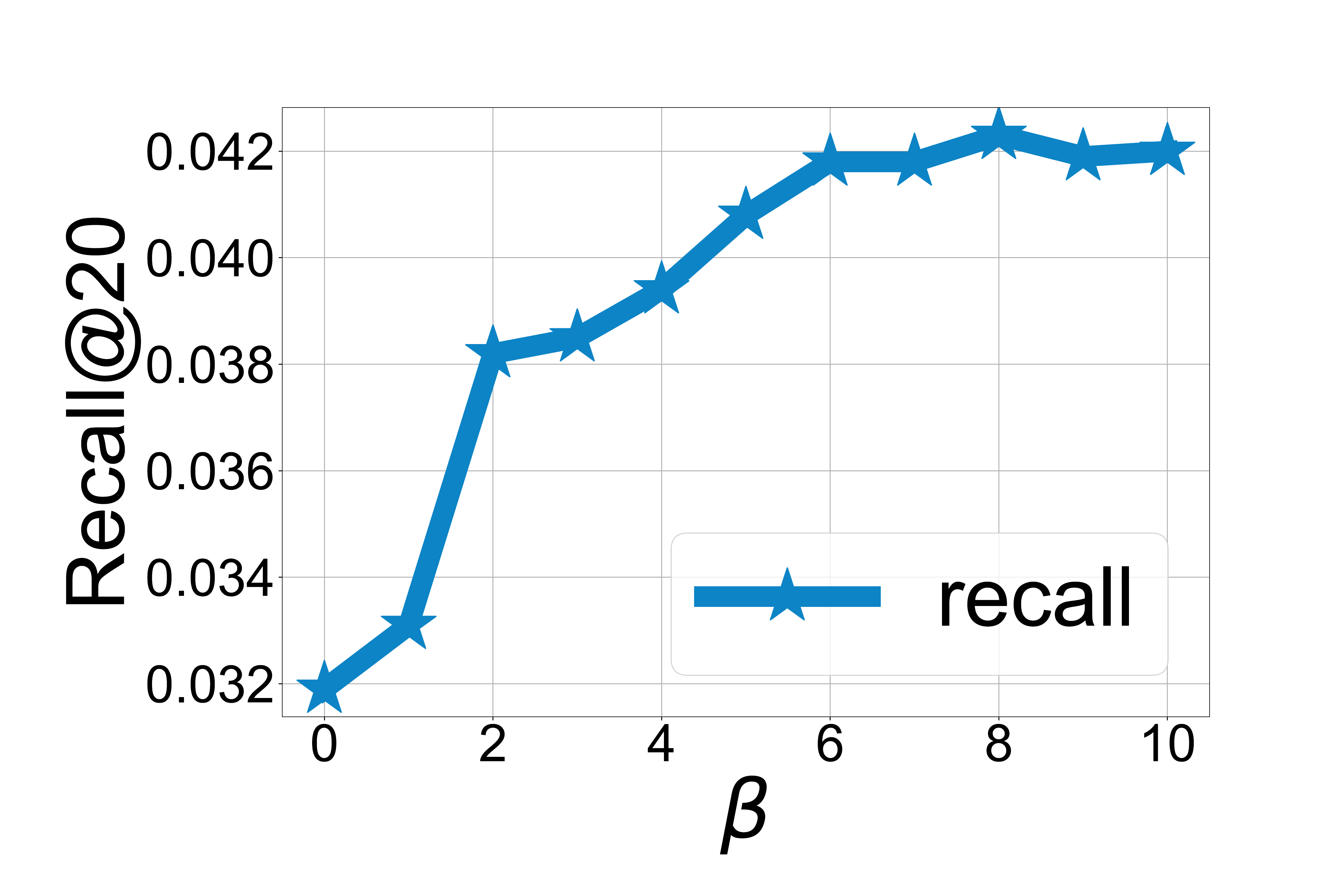}
    \caption{MF-Amazon}
  \end{subfigure}
  \begin{subfigure}[b]{0.23\textwidth}
    \centering
    \includegraphics[width=\linewidth]{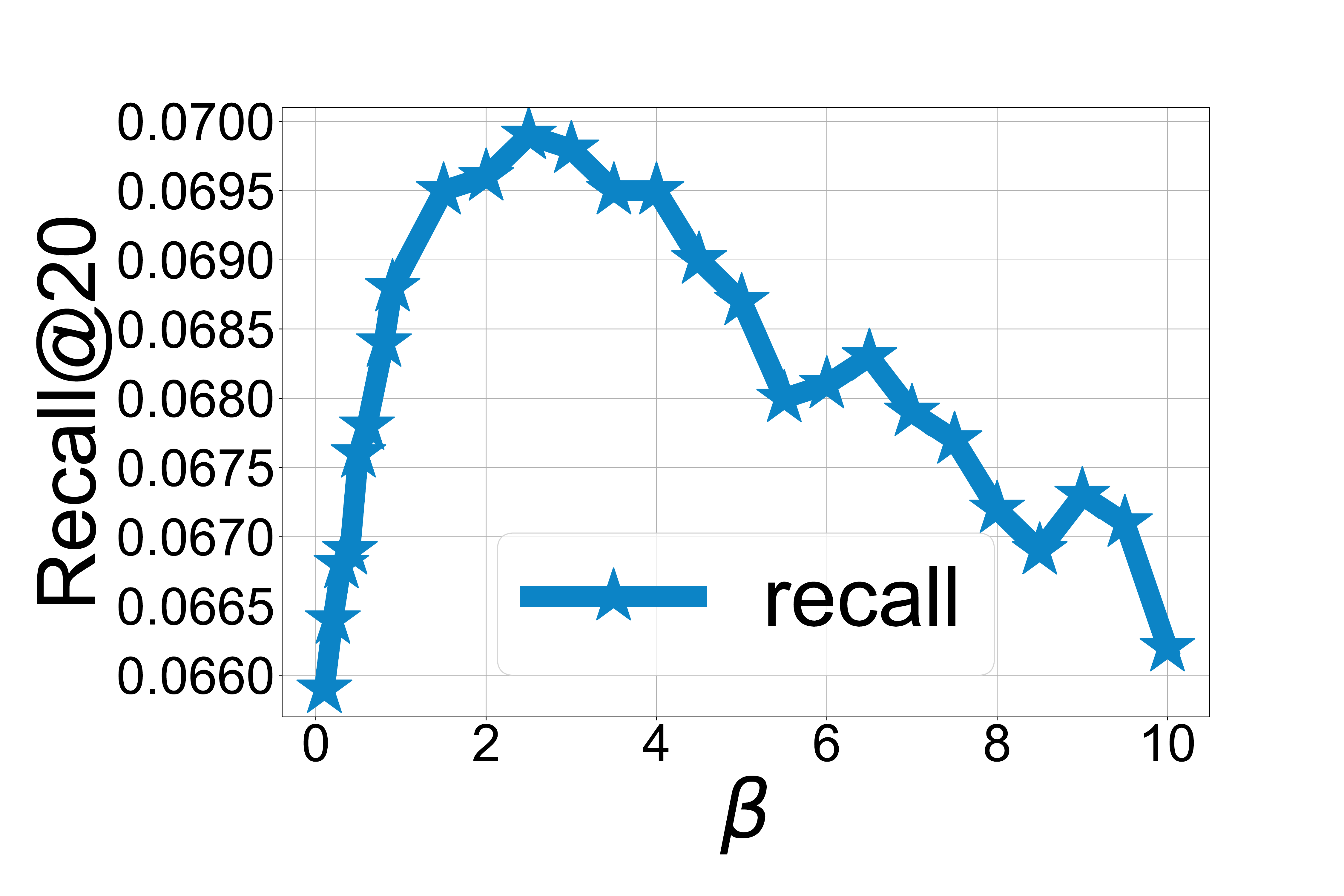}
    \caption{MF-Yelp2018}
  \end{subfigure}
% 这里是第一排的位置，全部放NDCG@20
  \begin{subfigure}[b]{0.23\textwidth}
    \centering
    \includegraphics[width=\linewidth]{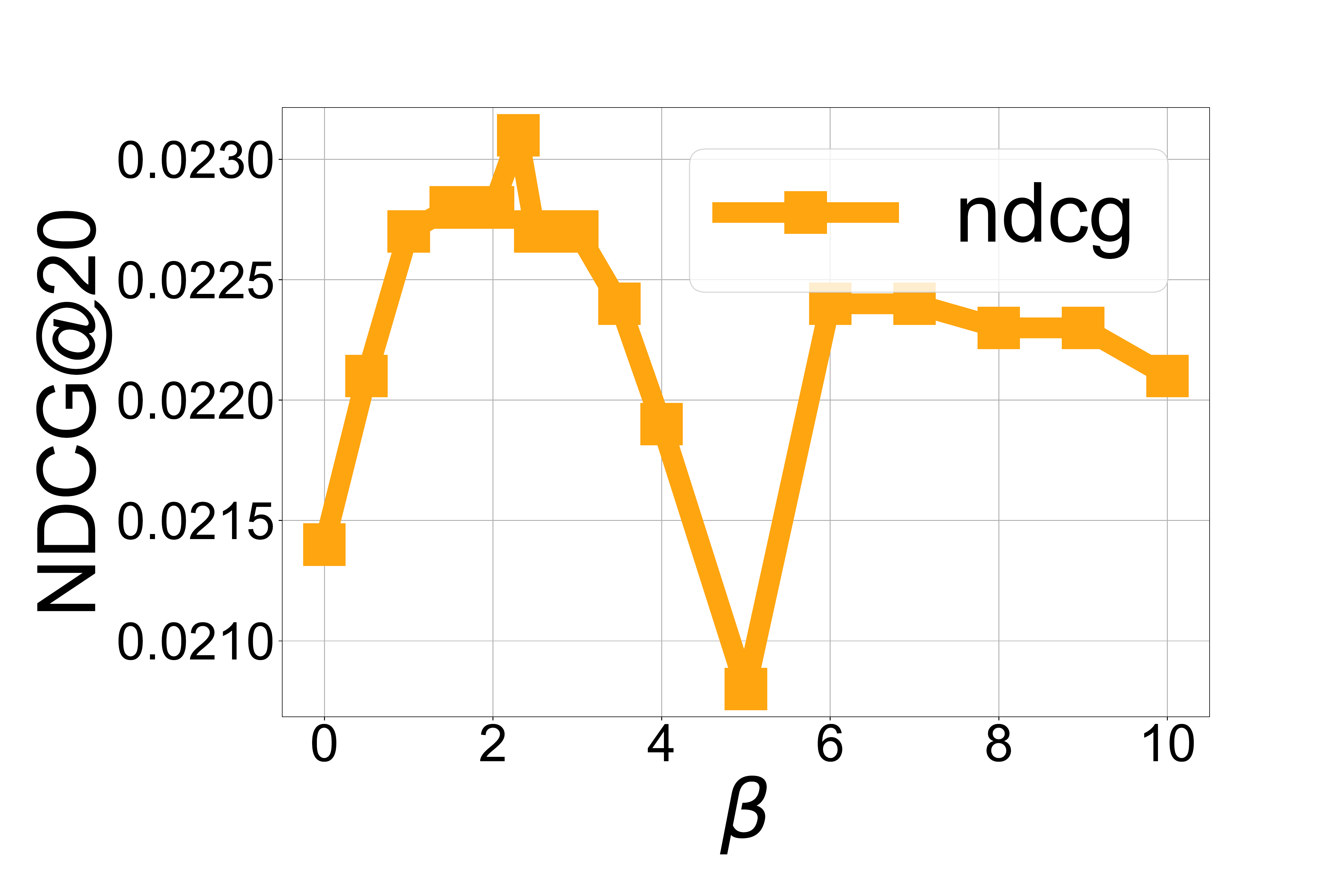}
    \caption{LightGCN-Amazon}
  \end{subfigure}
  \begin{subfigure}[b]{0.23\textwidth}
    \centering
    \includegraphics[width=\linewidth]{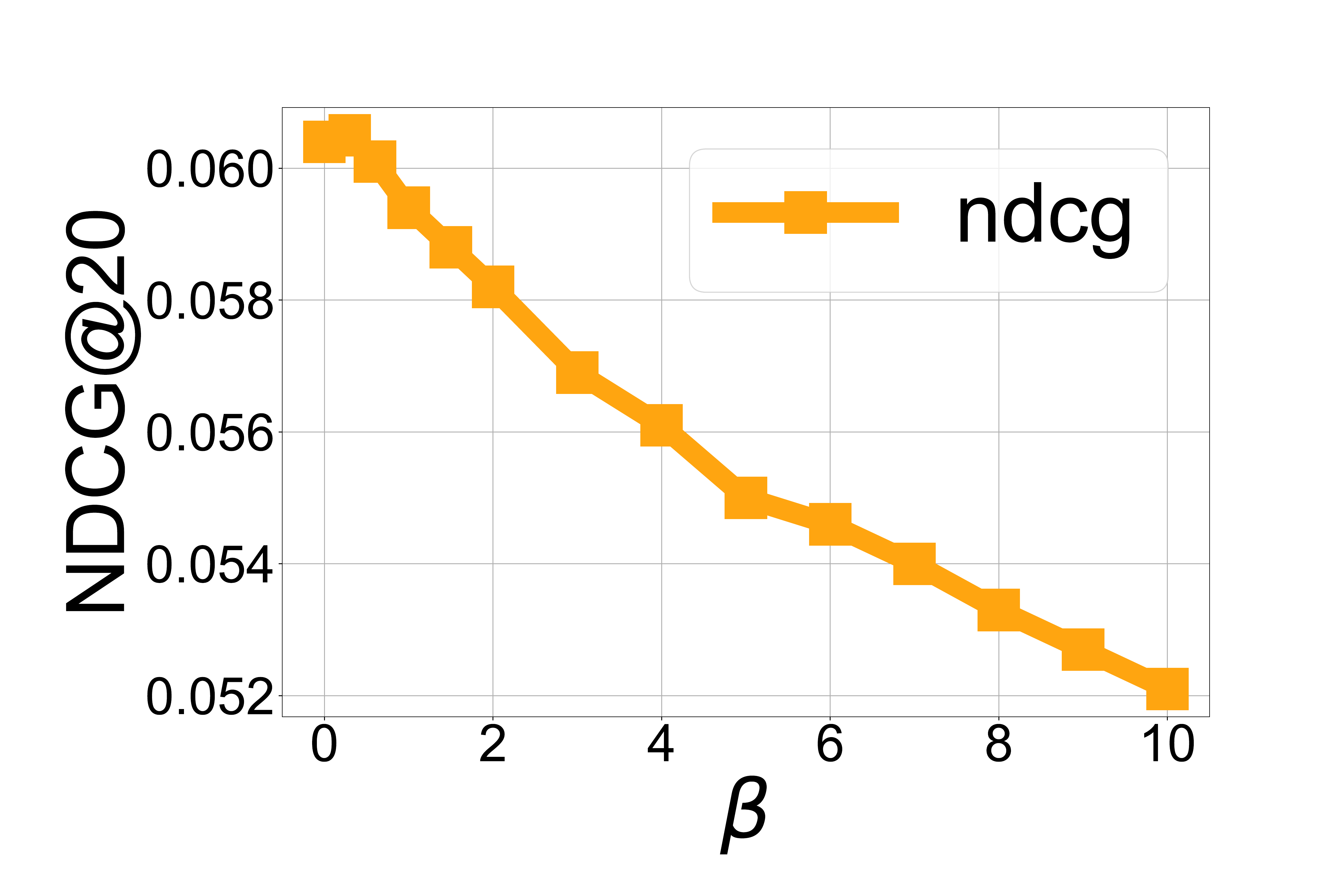}
    \caption{LightGCN-Yelp2018}
  \end{subfigure}
  % \caption{MF-Amazon}
  \begin{subfigure}[b]{0.23\textwidth}
    \centering
    \includegraphics[width=\linewidth]{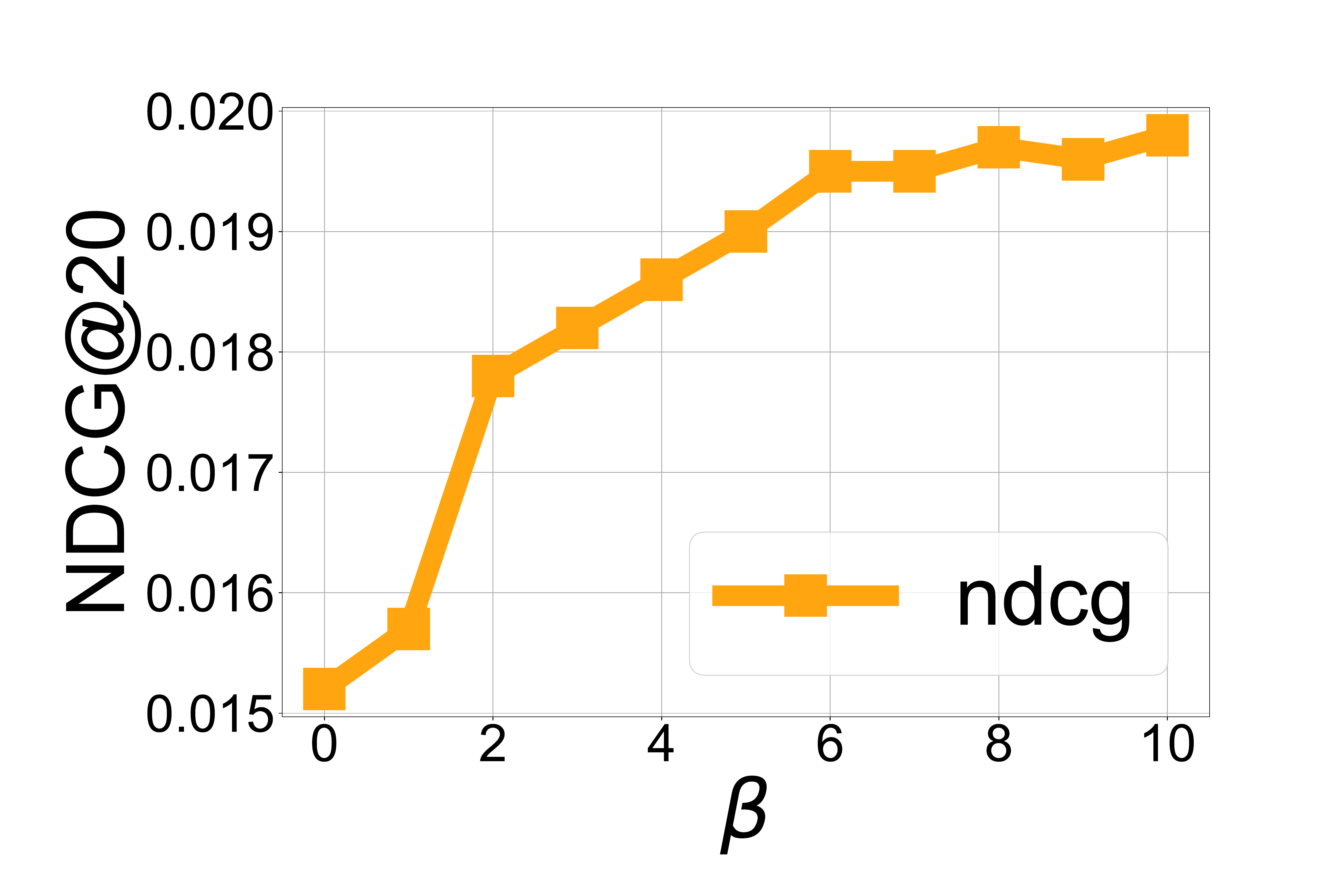}
    \caption{MF-Amazon}
  \end{subfigure}
  \begin{subfigure}[b]{0.23\textwidth}
    \centering
    \includegraphics[width=\linewidth]{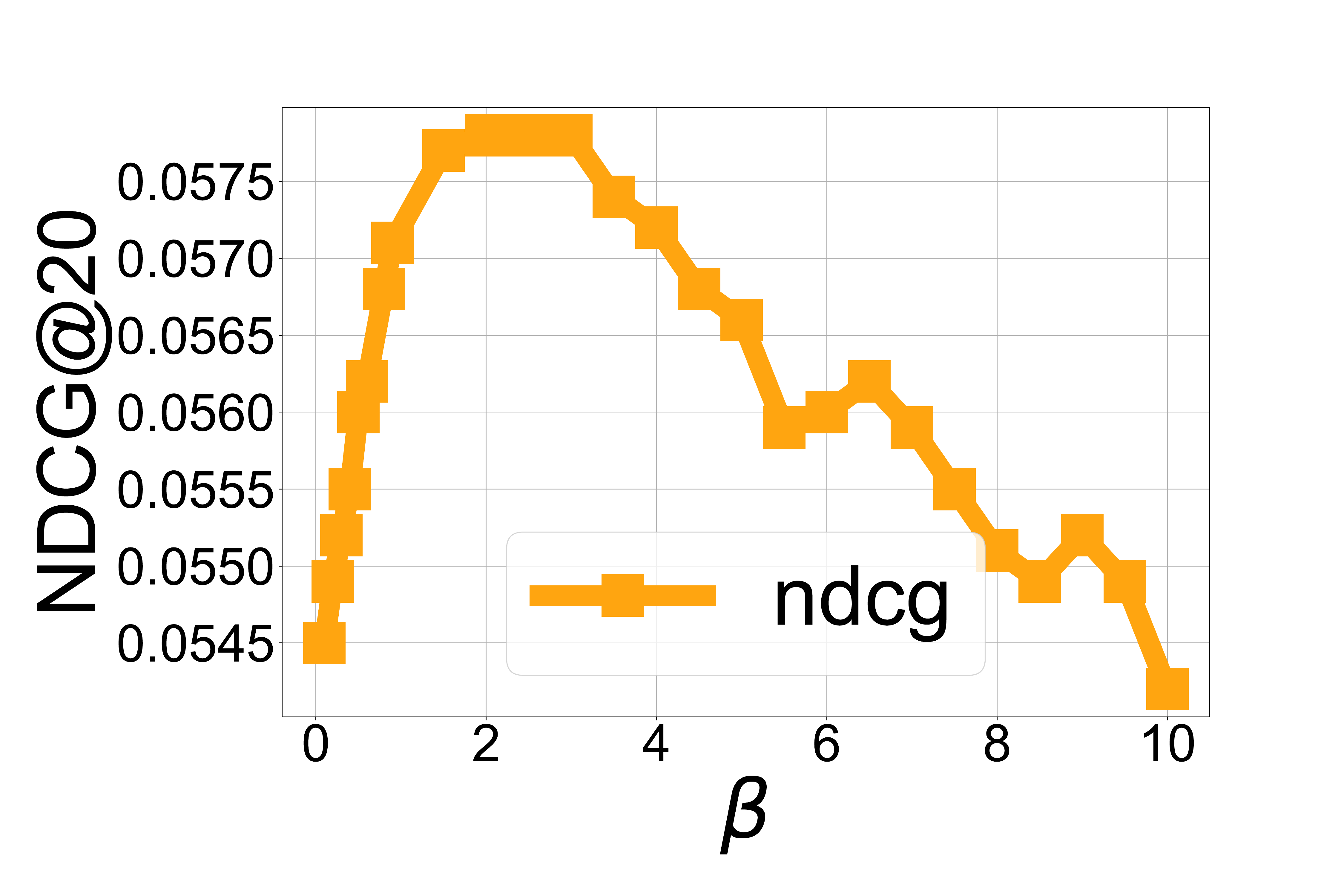}
    \caption{MF-Yelp2018}
  \end{subfigure}

  \caption{The impact of the $\beta$ value on model performance.}
  \label{fig:alpha value}
\end{figure*}

\subsection{RQ2: Ablation Study}
In this section, we conduct an ablation study of the three modules of~\modelname on LightGCN by building $3$ variants. A is built by removing the Hard Boundary Definition module, where we randomly sample a boundary item from the candidate set. B is built by replacing the dimension independent mixup module with the traditional mixup. C is built by removing the multi-hop pooling module by only considering the negative item for the first layer aggregation. Corresponding experimental results are presented in Figure~\ref{fig:ablation}. We can have the following observations:
\begin{itemize}[leftmargin=*]
    \item \modelname always performs the best. Removing any module would have a negative effect on the performance. It reveals each part of \modelname contributes to the performance.
    \item Removing the Hard Boundary Definition module (Variant A) results in a substantial drop in performance across the datasets, indicating the importance of maintaining an optimal size for the negative sampling region. A too-small region may lead to the loss of valuable feature information from the negative samples. At the same time, a too-large region hampers the effective fusion of information from the positive samples.
    \item Compared with the other two modules, changing the dimension independent mixup module to the traditional mixup (Variant B) has a relatively small performance drop. It shows the mixup technique is already powerful by generating negative items in the continuous space. The proposed dimension independent mixup further enhances the improvement, especially on the Alibaba.
    \item Removing the multi-hop pooling module has a great impact on the performance. It reveals the importance of considering high-order information during the negative sampling procedure. This observation aligns with the finding of MixGCF~\cite{Huang2021MixGCF}.
\end{itemize}

By observing the data, it can be seen that the area of negative sampling can neither be too large nor too small. According to Equation \ref{equation:mix}, if the area is too small, the hard negative item is synthesized too close to the positive items and thus the features of the negatives will be lost, while if the area is too far, the features of the positive items cannot be incorporated into the hard negative items at all.

\begin{figure*}[t]
  \centering
  % 第二版画法，占两栏，一行四个
  \begin{subfigure}[b]{0.23\textwidth}
    \centering
    \includegraphics[width=\linewidth]{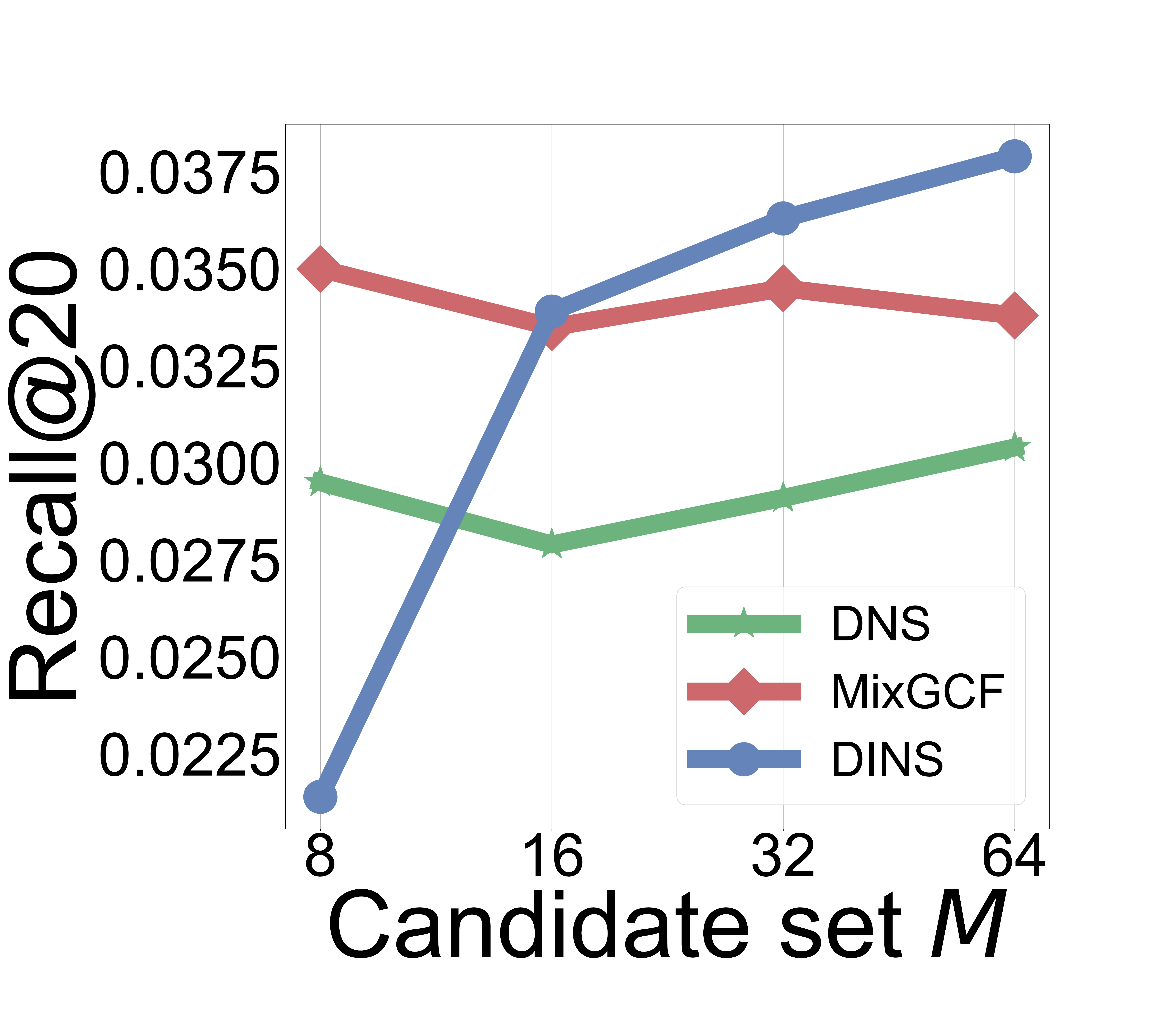}
    \caption{NGCF-Amazon}
  \end{subfigure}
  \begin{subfigure}[b]{0.23\textwidth}
    \centering
    \includegraphics[width=\linewidth]{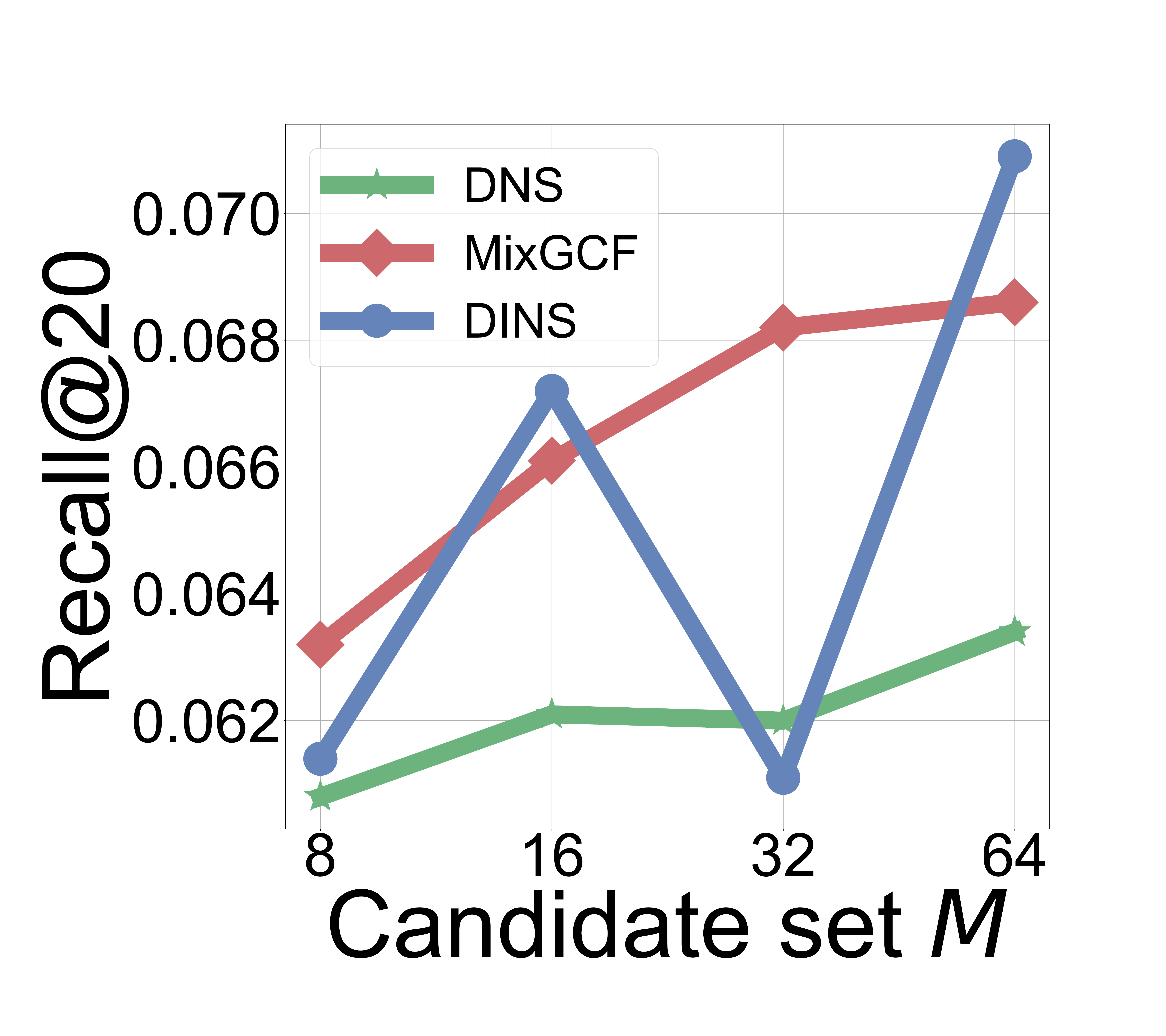}
    \caption{NGCF-Yelp2018}
  \end{subfigure}
  % \caption{NGCF-Amazon}
  \begin{subfigure}[b]{0.23\textwidth}
    \centering
    \includegraphics[width=\linewidth]{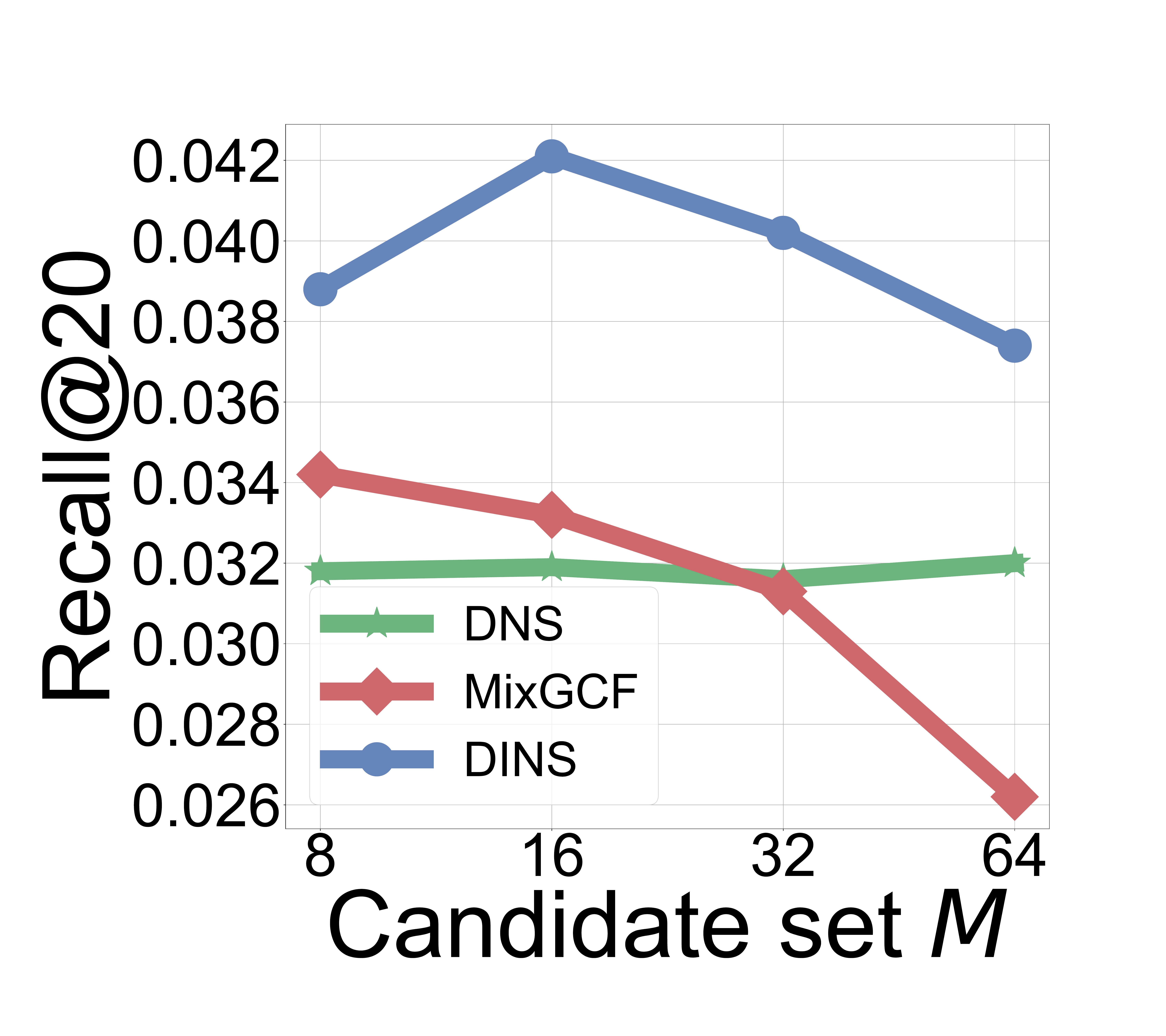}
    \caption{MF-Amazon}
  \end{subfigure}
  \begin{subfigure}[b]{0.23\textwidth}
    \centering
    \includegraphics[width=\linewidth]{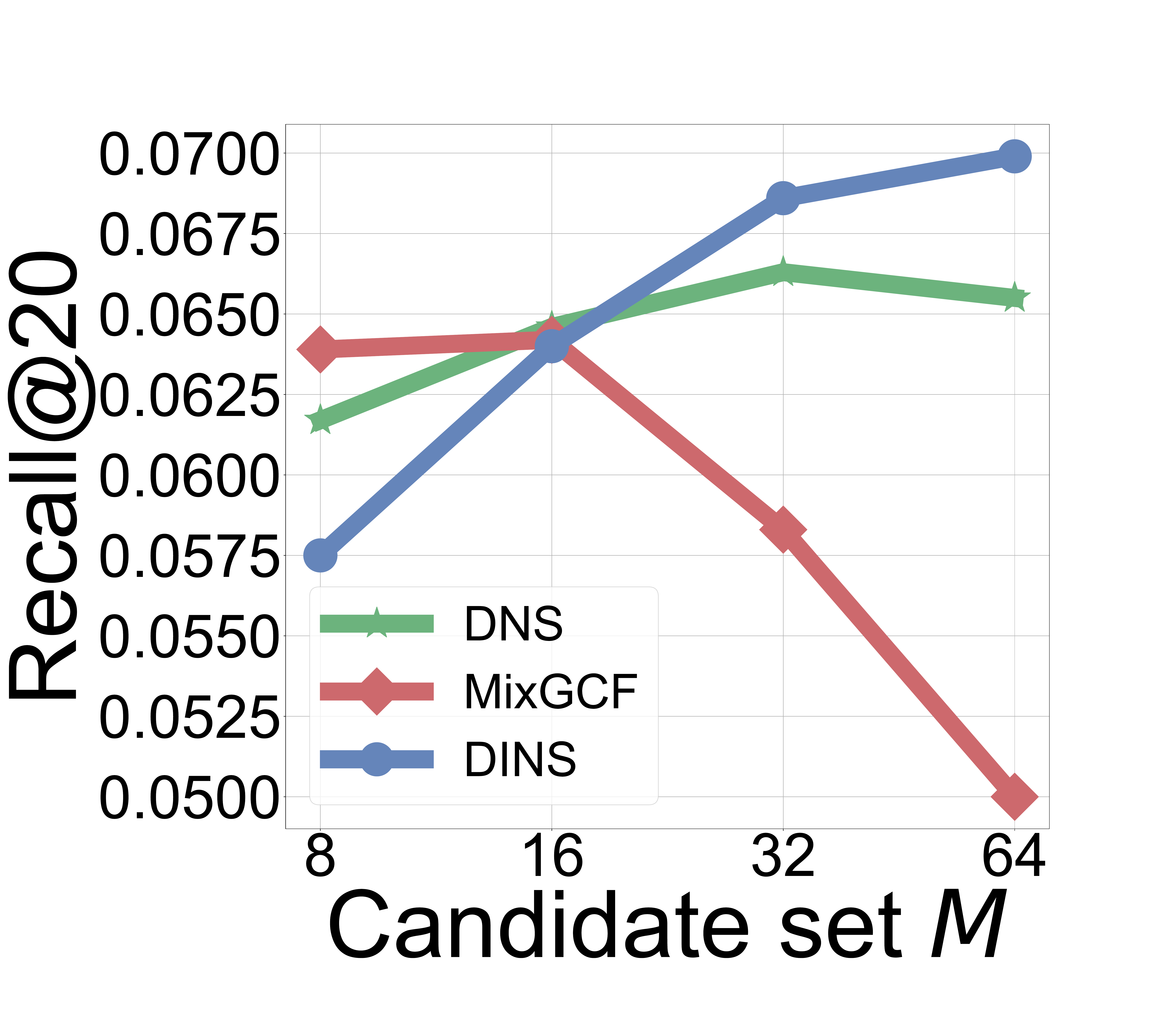}
    \caption{MF-Yelp2018}
  \end{subfigure}

  \begin{subfigure}[b]{0.23\textwidth}
    \centering
    \includegraphics[width=\linewidth]{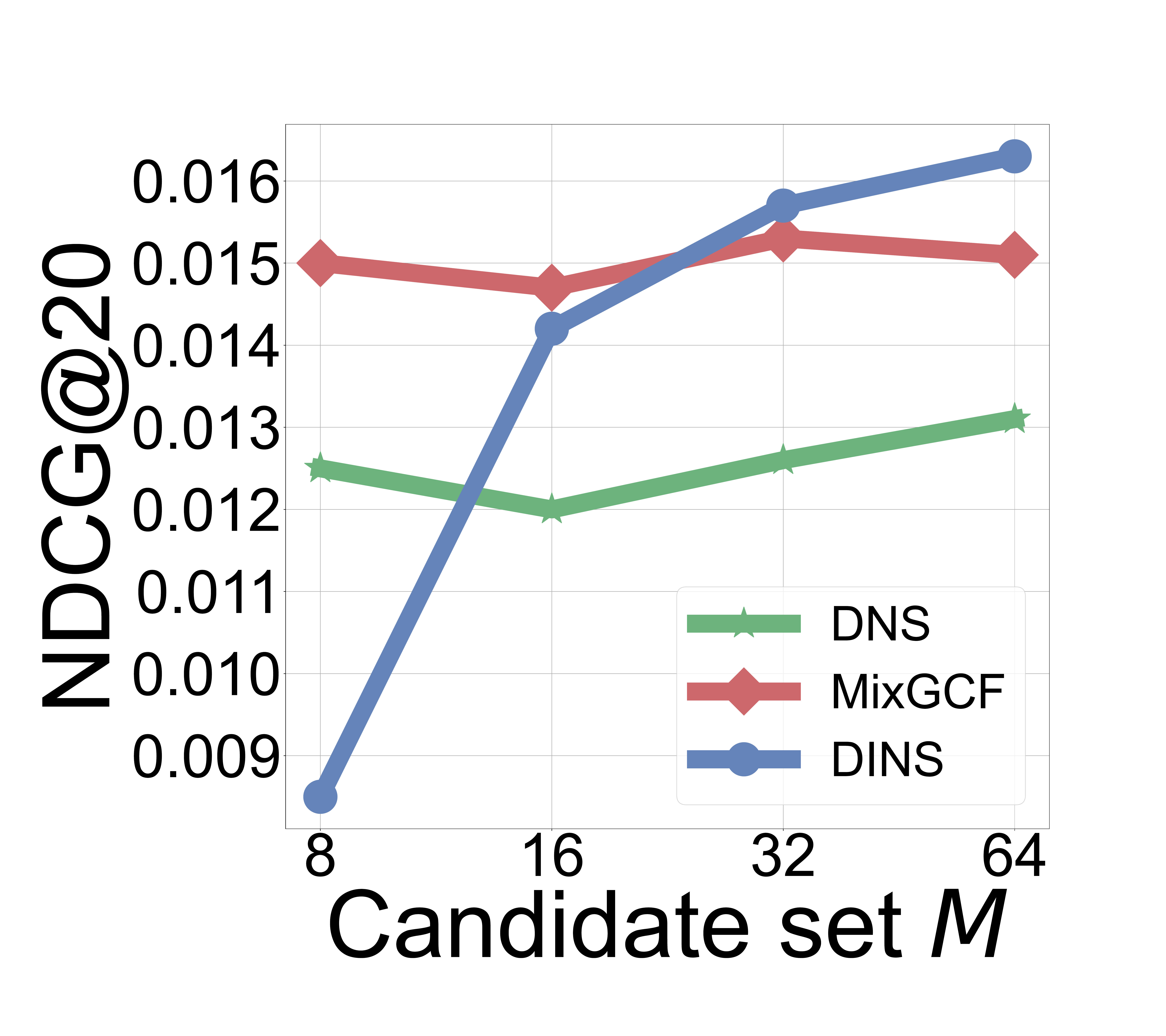}
    \caption{NGCF-Amazon}
  \end{subfigure}
  \begin{subfigure}[b]{0.23\textwidth}
    \centering
    \includegraphics[width=\linewidth]{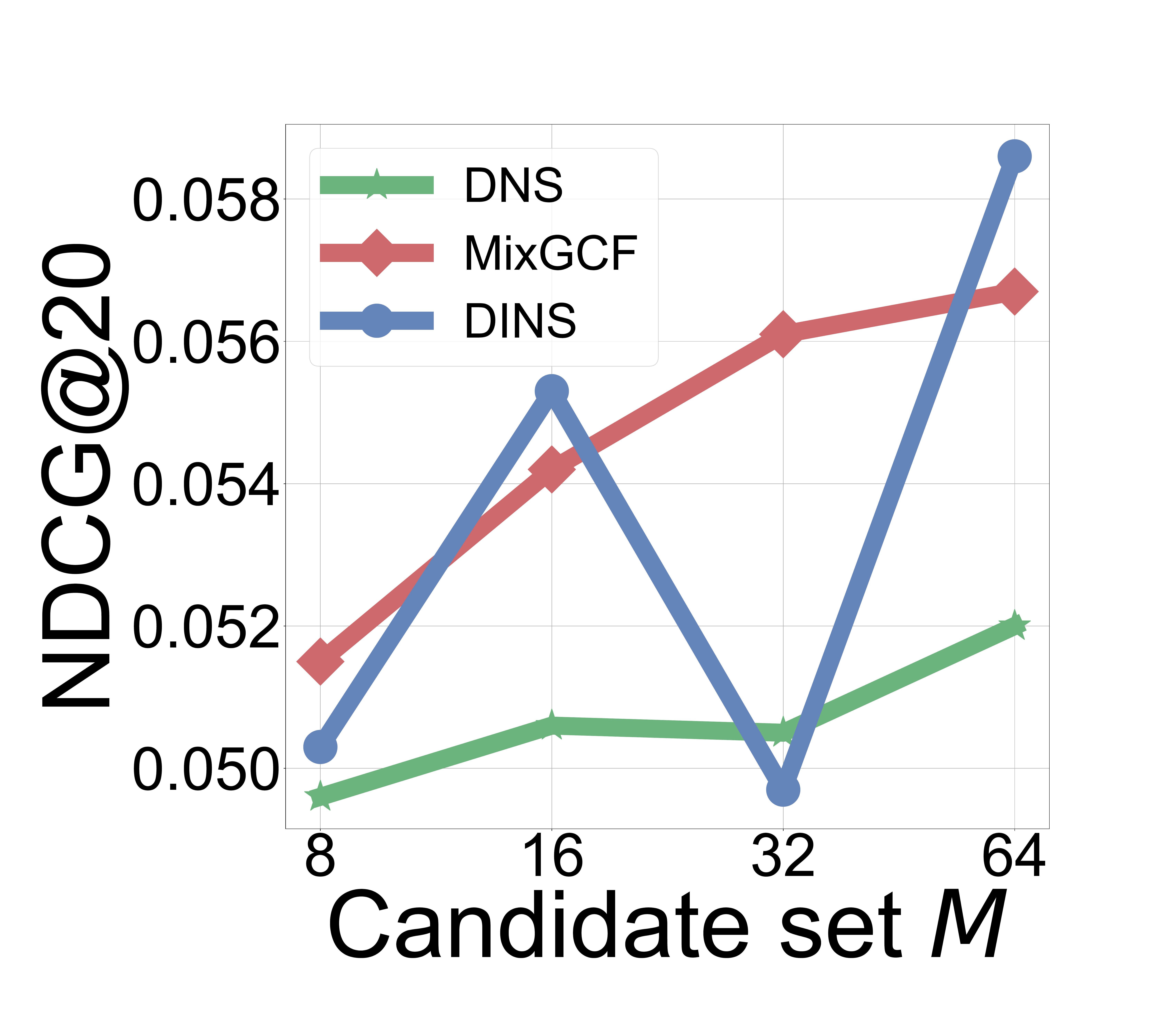}
    \caption{NGCF-Yelp2018}
  \end{subfigure}
  % \caption{NGCF-Amazon}
  \begin{subfigure}[b]{0.23\textwidth}
    \centering
    \includegraphics[width=\linewidth]{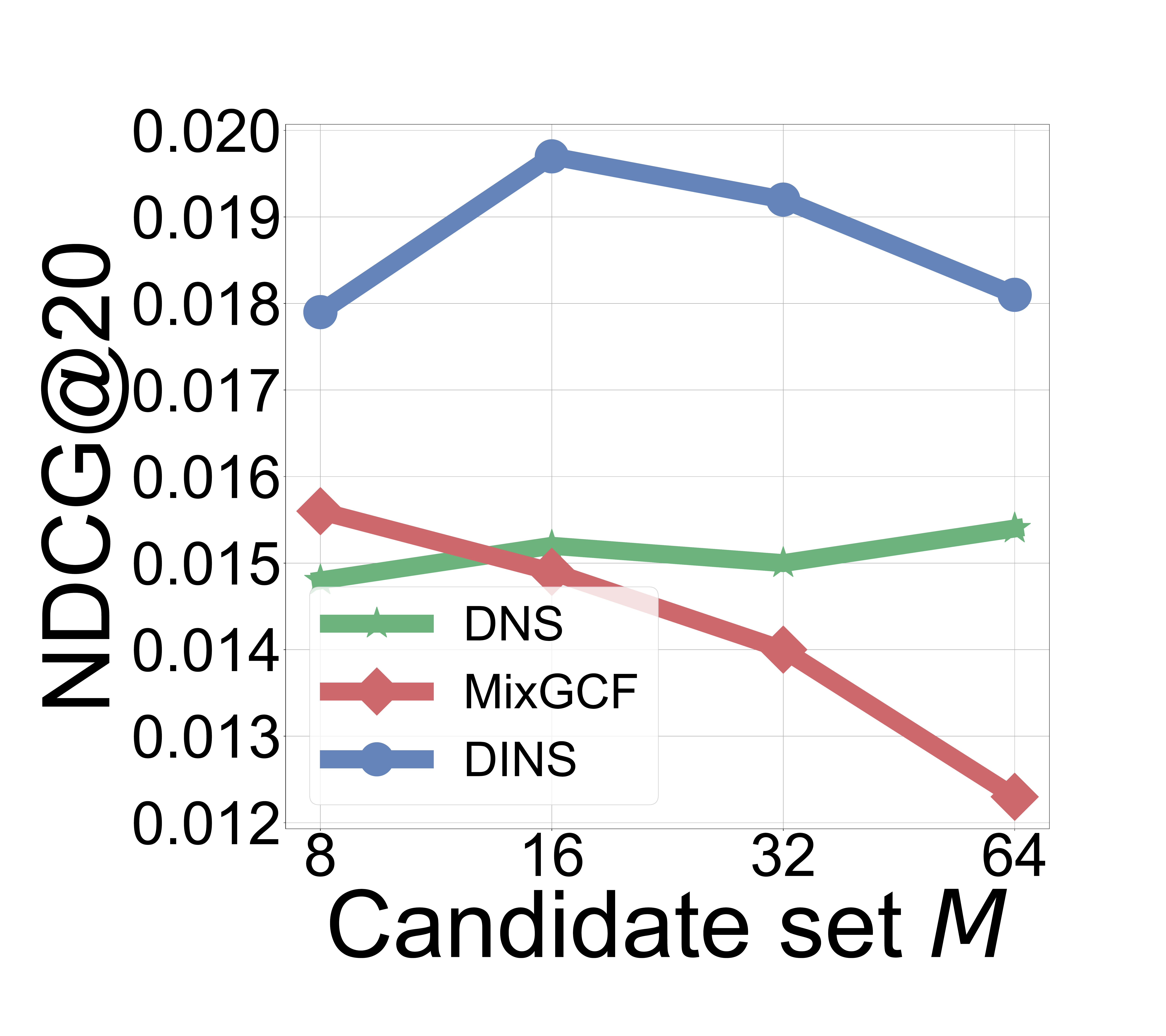}
    \caption{MF-Amazon}
  \end{subfigure}
  \begin{subfigure}[b]{0.23\textwidth}
    \centering
    \includegraphics[width=\linewidth]{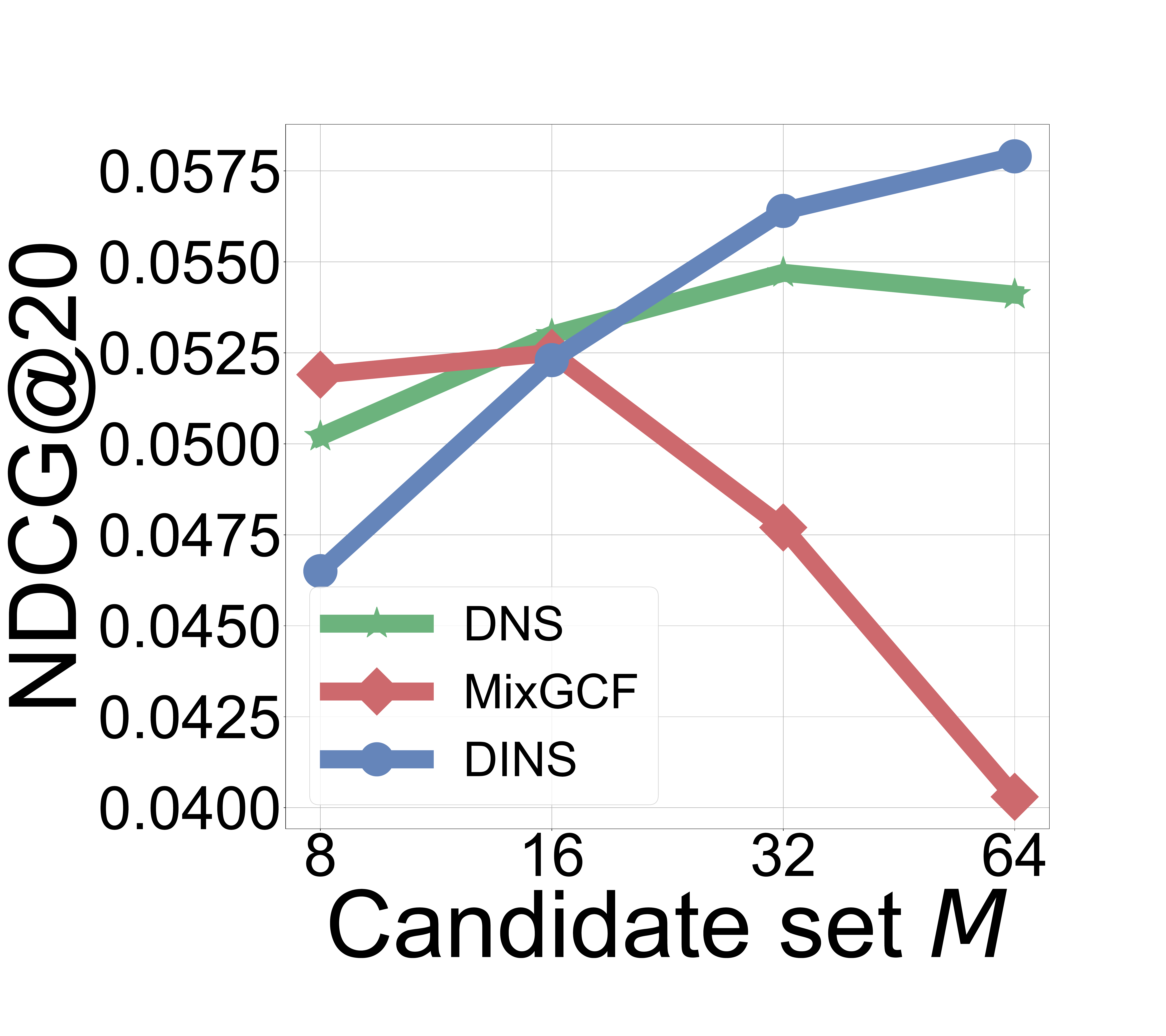}
    \caption{MF-Yelp2018}
  \end{subfigure}

  \caption{The impact of the candidate set size $M$.}
  \label{fig:candidate set}
\end{figure*}

\subsection{RQ3: Parameter Sensitivity}
In this section, we focus on how the hyper-parameters ($\beta$, $M$) and the boundary item selection method actually affect \modelname.

\subsubsection{Impact of the $\beta$ value} $\beta$ is a distinctive hyper-parameter within our \modelname framework, serving to regulate the incorporation of information from the positive item as a collective entity into our synthetic hard negative item. $\beta$ is a propensity coefficient that controls whether our synthetic \textbf{Hard Negative Samples} (HNS) as a whole converge toward positive or negative items. Larger $\beta$ means the synthetic HNS is closer to the positive item. We conduct experiments with all $3$ datasets on $3$ recommenders. Due to the space limit, we illustrate parts of the results in Figure~\ref{fig:alpha value}. Similar trends are also observed in other experiments. 
\begin{itemize}[leftmargin=*]
    \item As the dataset becomes sparser, the encoder necessitates the fusion of a greater amount of information about the positive items. According to the statistics in Table~\ref{table1}, it is clear that Yelp2018 is much denser than Amazon and Alibaba. On the Yelp2018 dataset, LightGCN achieved the best results at around $\beta = 0.1$ and MF at around $\beta = 2$. In contrast, on the Amazon dataset, LightGCN needs to be around 2.3 and MF needs to be around 8 to achieve the best results. This may be due to the fact that the sparser the dataset the more uniform the embedding distribution of users and items is, resulting in a larger sampling boundary for determination. Even after dimension-wise mixup, the synthesized negative items can still be far away from the positive one, so we need to move closer to the positive item as a whole with a larger bet.
    \item The stronger the encoding capability of the encoder, the less information from the positive item is needed for \modelname. LightGCN is a state-of-the-art CF method that employs a message-passing mechanism on the bipartite user-item graph to learn a better embedding and MF is a straightforward model which recommends by a factorization-based approach. The $\beta$-values required for LightGCN to achieve the best performance on all three datasets are much smaller than those of MF. This may be because LightGCN uses a message-passing mechanism on the graph to automatically aggregate a portion of the information from the positive items, while MF does not have a message-passing mechanism to naturally aggregate the information from the positives. 
\end{itemize}
   \subsubsection{Impact of the candidate set size $M$} We also conducted an experimental analysis on the candidate set size for negative items in DNS and MixGCF. Then we tested these baselines using candidate set sizes of 8, 16, 32, and 64. Detail experiments are illustrated in~Figure~\ref{fig:candidate set}.
We can observe that the $M$ impacts all the sampling methods.
Generally, increasing the candidate set size $M$ tends to improve the performance of the experiment. For example, the best results are mostly achieved with $M=64$.
Interestingly, MixGCF exhibits lower stability than DNS and \modelname, even yielding contrasting outcomes on the Amazon dataset. Notably, as the candidate set increases, MixGCF demonstrates a significant decline. We attribute this disparity in results to the attributes of the dataset.

\subsubsection{Impact of the boundary item selection method.}
To delve deeper into the influence of the boundary item selection method in Section~\ref{sec:boundary}, we further conduct experiments on the impact of the selection method. In the experiment, we opt for LightGCN as the encoder due to its superior performance. To be noted that the sampled boundary item decides the sampling area together with the positive item. The area is the continuous space between the boundary and the positive item. By multiplying all the absolute differences in each dimension, we can obtain the volume of the sampling area. Based on this principle, we further design $3$ sampling methods to find the boundary item as:
% 我们设计了以下5种计算采样面积的方法
\begin{itemize}[leftmargin=*]
    \item \textbf{Random}. Randomly select an item from the candidate set as the boundary item.
    \item \textbf{Min}. Find the item that constitutes the minimum sampling area volume as the boundary item.
    \item \textbf{Max}. Find the item that constitutes the maximum sampling area volume as the boundary item.
    \item \textbf{Dot Product (DP)}. The method used in \modelname as Equation~\ref{equation:HNS}.
\end{itemize}
After obtaining the boundary item, we also conduct the dimension independent mixup and multi-hop pooling. Subsequently, we conducted experiments on LightGCN to examine the impact of the boundary item selection method across the three datasets. The detailed experimental results are shown in Table \ref{table:area_boundary_sampling}. We can observe that Min and Max perform badly in most cases, which reveals the importance of selecting a suitable area for generating negative items. Even random selection outperforms Min and Max methods. The method used by \modelname (DP) always performs the best across the $3$ datasets. It shows we have selected a suitable sampling area to generate the negative item.

\begin{table}[]
\caption{Performance Comparison of different boundary item selection methods.}\label{table:area_boundary_sampling}
\begin{tabular}{c|cc|cc|cc}
\toprule
                             & \multicolumn{2}{c}{Amazon}  & \multicolumn{2}{c}{Alibaba} & \multicolumn{2}{c}{Yelp2018} \\ \cline{2-3} \cline{4-5} \cline{6-7}
                            & R@20       &  {N@20}         & R@20      &  {N@20}         & R@20     & N@20 \\ \hline
Random                        & 0.0300       & 0.0123     & 0.0460        & 0.0201       & 0.0653        & 0.0537            \\
Min                          & 0.0145       & 0.0055     & 0.0279        & 0.0129       & 0.0441        & 0.0360              \\
Max                          & 0.0246       & 0.0084     & 0.0182        & 0.0081       & 0.0357        & 0.0294             \\ \hline DP                  & \textbf{0.0493}       & \textbf{0.0231}    & \textbf{0.0764}        & \textbf{0.0358}       & \textbf{0.0738}        & \textbf{0.0606}              \\ \bottomrule
% DP*               & 0.0477       & 0.0225  & 0.0726        & 0.0348       & 0.0731        & 0.0598             \\ \bottomrule
\end{tabular}
\end{table}

\subsection{RQ4: Case Study}
To answer RQ4: \textit{Does \modelname really support area-wise hard negative sampling?} we conduct a case study for readers to understand the sampling principle behind \modelname. Experiments are conducted on the Yelp2018 dataset with LightGCN as the backbone recommender. We train the model with RNS, MixGCF, and \modelname for $60$ epochs. For a fixed user-item interaction, we store the embedding of the positive item and the sampled negative item by different sampling methods in each iteration. Then we obtain the averaged positive item representation by mean pooling over all the collected positive item embedding. Then we concatenate the averaged positive item and all collected negative items embedding together and visualize the distribution via t-SNE~\cite{van2008visualizing}. For better visualization, we move the positive item to the center of the visualization by subtracting all the reduced $2$-d embedding from the reduced positive item embedding.

\begin{figure}
  \centering
  \begin{subfigure}[b]{0.155\textwidth}
    \centering
    \includegraphics[width=\linewidth]{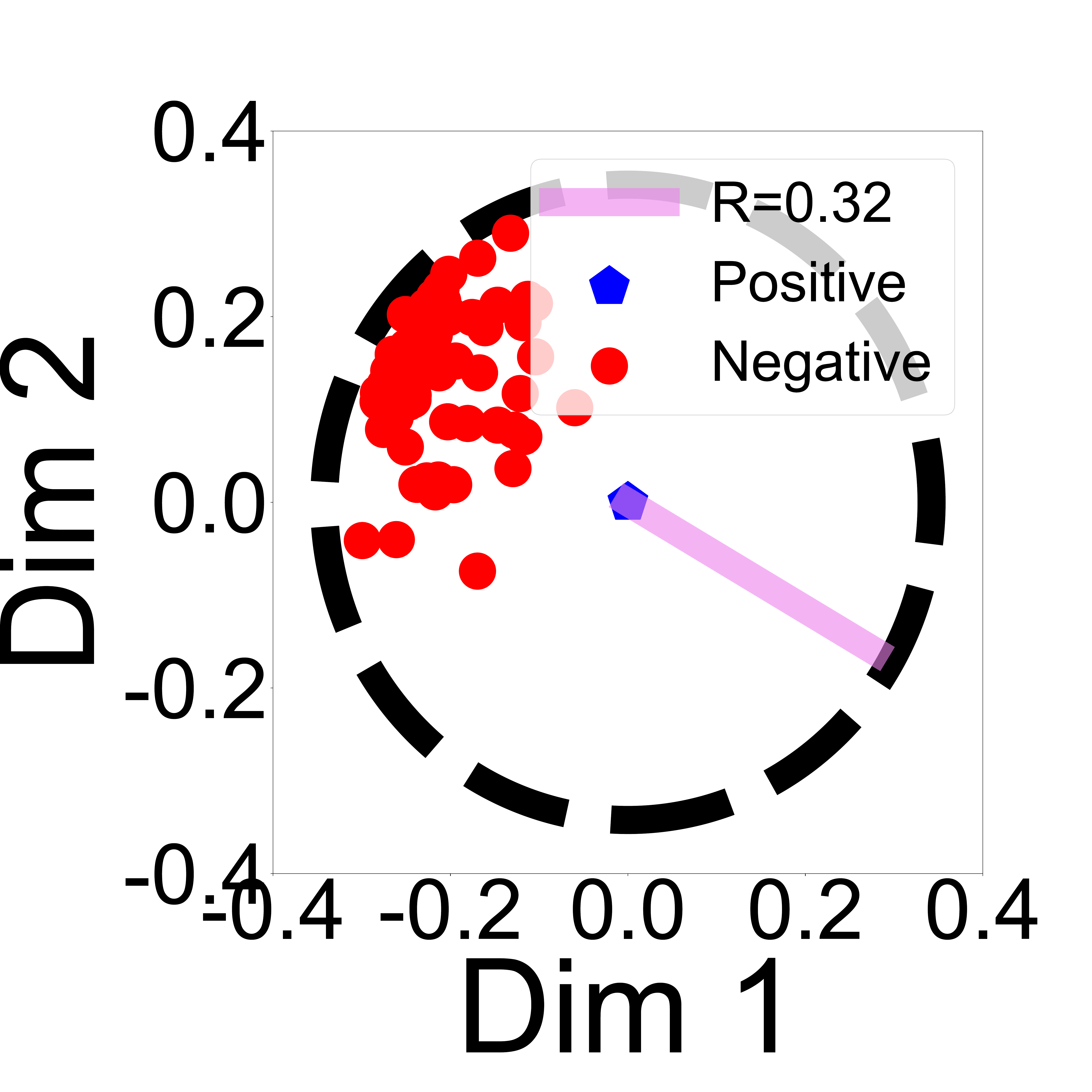}
    \caption{RNS sampling}
    \label{fig:subfig1}
  \end{subfigure}
  \begin{subfigure}[b]{0.155\textwidth}
    \centering
    \includegraphics[width=\linewidth]{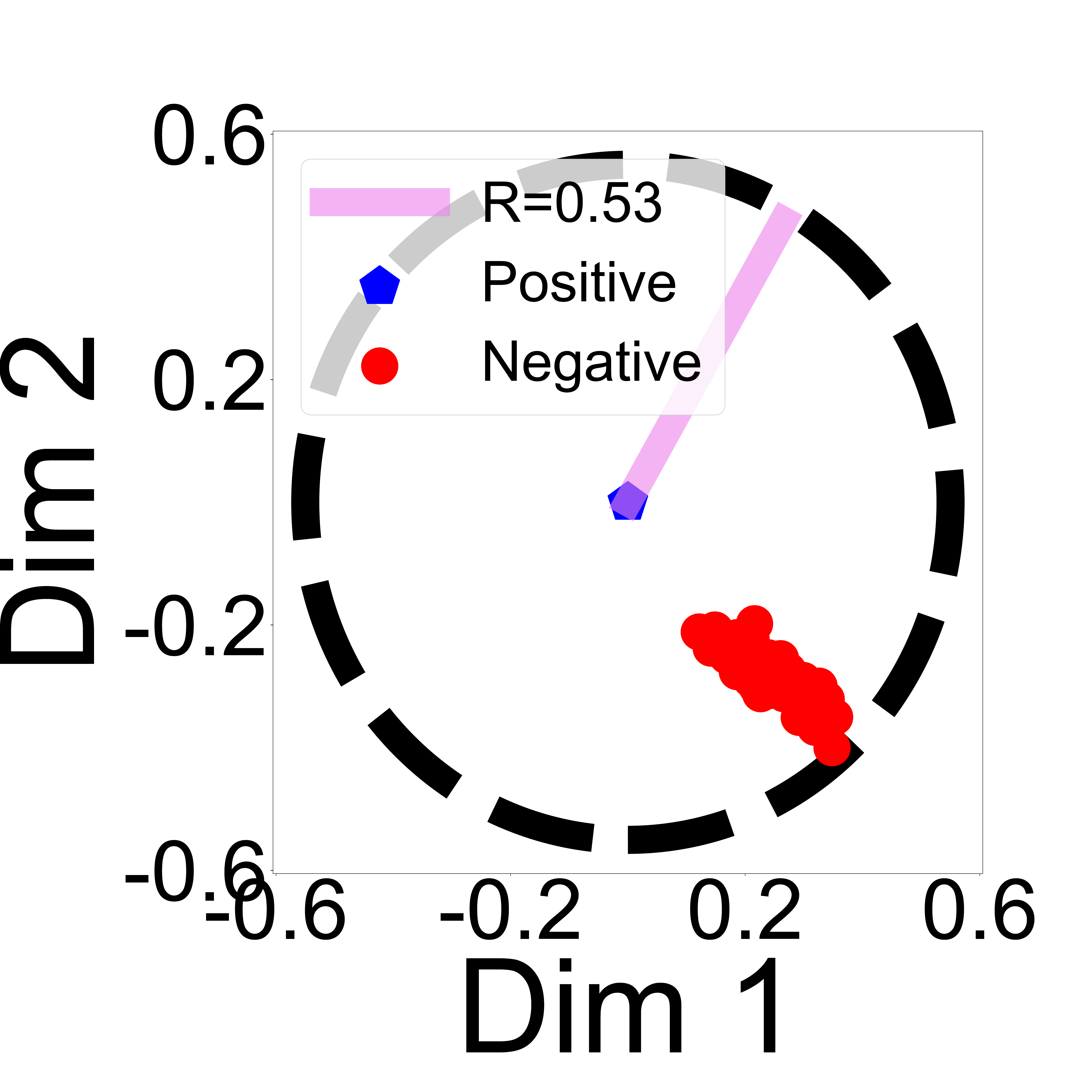}
    \caption{MixGCF sampling}
  \end{subfigure}
  \begin{subfigure}[b]{0.155\textwidth}
    \centering
    \includegraphics[width=\linewidth]{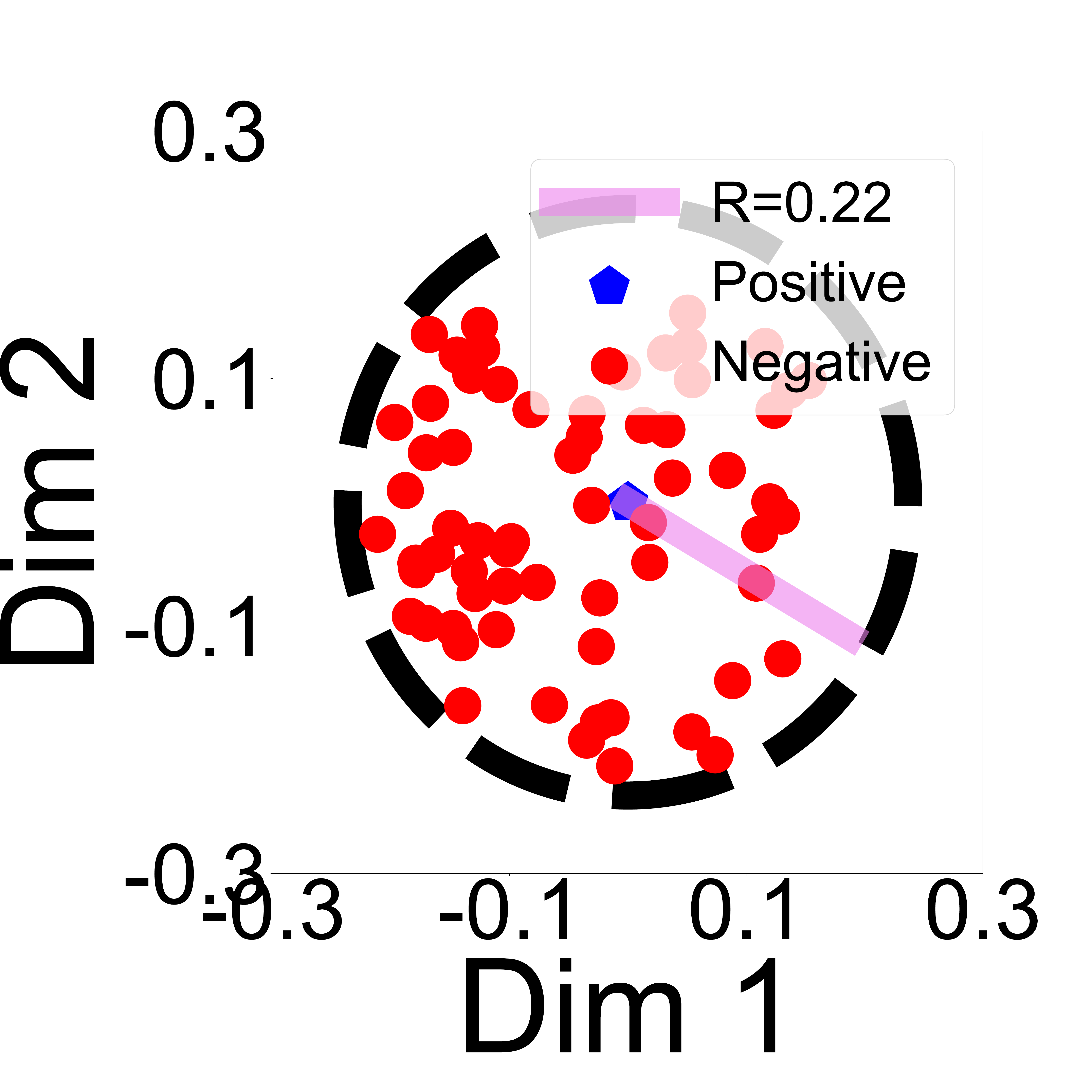}
    \caption{\modelname sampling}
  \end{subfigure}
    \caption{The hard negatives sampled by RNS, MixGCF,\modelname}
    \label{fig:negative_compare}
\end{figure}

The visualization is shown in Figure~\ref{fig:negative_compare}. We draw a circle (with a radius represented by $R$) to show the farthest sampled negative item. We can have three observations:
\begin{itemize}[leftmargin=*]
    \item The RNS sampling method samples varied negative items that are far from the positive item. It exhibits the characteristic of the point-wise negative sampling method. It samples other existing items that are restricted in the upper-left corner area based on currently learned embedding. 
    \item The MixGCF method, as expected for the line-wise sampling method, exhibits line-style negative sampling results. It is due to the traditional Mixup method that generates a linear interpolation of two embeddings with the same weight on all dimensions.
    \item \modelname shows the characteristics of the area-wise sampling method. We can observe the sampled negative items spans across the whole circle, which gives a sufficient exploration of the embedding space. At the same time, the sample radius of \modelname is only $0.22$, which is the smallest among the three methods. It shows \modelname samples the hard negatives for assist model training.
\end{itemize}

%%%     这是三种负采样方式采样出来的负样本分布图

\section{Related Work}\label{sec:related_work}
In this section, we introduce the related work for \modelname, which includes the graph-based recommendation and negative sampling method in recommendation.
\subsection{Graph-based recommendation}
Recent years have witnessed the rapid development of the emerging direction of GNN-based recommender systems~\cite{He2020lightgcn, SVD_GCN,Wang2019NGCF, pinsage, GC_MC,yang2021consisrec,yang2022large,wang2022metakrec,yang2023dgrec} where the user-item interactions are presented as a bipartite graph, and graph neural network methods are employed to learn the representation of each node via exploring structure information. 
For example, Pinsage~\cite{pinsage} samples neighborhoods according to the visit counts of a node through a random-walk sampling approach. 
GC-MC~\cite{GC_MC} proposes a graph auto-encoder framework to construct the node representation by directly aggregating the information of its neighbors.
NGCF~\cite{Wang2019NGCF} captures high-order connectivities by stacking multiple embedding propagation layers and utilizes the combination of different layers' output for the rating prediction.
Compared with NGCF, LightGCN~\cite{He2020lightgcn} achieves better training efficiency and generation capability by removing the feature transformation and nonlinear activation function.
Finally, SVD-GCN~\cite{SVD_GCN} further simplifies LightGCN by replacing neighborhood aggregation with exploiting K-largest singular vectors for the close relation between GCN-based and low-rank methods.
\subsection{Negative sampling method}
\label{negative sampling}
Negative sampling methods in RecSys have gained significant attention due to their ability to accelerate training and greatly enhance model performance, which can be categorized into the following groups.
\textbf{Static Sampler} usually selects negatives from items that the user has not interacted with yet, based on a pre-defined distribution, like uniformity distribution~\cite{He2020lightgcn,Wang2019NGCF,Rendle2012BPR} or popularity distribution~\cite{pop-based}.
\textbf{Hard Negative Sampler} techniques choose negative items with the highest scores from the current recommender~\cite{DNS,SoftBPR}. There are some mixup-based methods~\cite{Huang2021MixGCF,DENS} generate new negative items by performing mix operations. 
\textbf{GAN-based methods} like IRGAN\cite{Wang2017IRGAN:} and AdvIR\cite{AdvIR} use adversarial learning to generate negative items and improve robustness. 
\textbf{Auxiliary-based Samplers} leverage additional information, such as the knowledge graph in KGPolicy\cite{KGPolicy20} or Personalized PageRank scores in PinSage\cite{pinsage}, to sample hard negative instances.

\section{Conclusion}\label{sec:conclusion}
In conclusion, this paper provides a novel perspective to revisit the current negative sampling methods based on continuous sampling area and classifies them into point-wise and line-wise sampling methods. In a further step, we design the first area-wise sampling method, named \modelname, by proposing the Dimension Independent Mixup method. \modelname can easily support both matrix factorization and graph-based backbone recommenders. Extensive experiments demonstrate superior performance compared with other methods, making it a state-of-the-art solution for negative sampling when training collaborative filtering with implicit feedback. The contributions of this work include a fresh perspective on negative sampling methods, the introduction of Area-wise sampling, and the development of the innovative \modelname method. These findings have the potential to enhance RecSys capabilities and improve user experiences in various online services.

\begin{acks}
This work is supported by Hebei Natural Science Foundation of China Grant F2022203072, CCF-Zhipu AI Large Model Fund (CCF-Zhipu202307), Innovation Capability Improvement Plan Project of Hebei Province (22567626H), supported in part by NSF under grants III-1763325, III-1909323,  III-2106758, and SaTC-1930941.
\end{acks}

% \newpage
\bibliographystyle{ACM-Reference-Format}
\bibliography{sample-base}

\appendix

\end{document}